\documentclass{aa}  

\usepackage{graphicx}
\usepackage{txfonts}

\usepackage{lipsum}
\usepackage{xcolor}
\usepackage{hyperref}
\usepackage[normalem]{ulem}
\usepackage{subcaption}
\usepackage{booktabs}
\usepackage{siunitx}

\definecolor{frcolor}{rgb}{0,0.5,0}

\newcommand{\msun}{M_\odot}

\newcommand{\mch}{M_\mathrm{Ch}}

\newcommand{\rhoini}{\ensuremath{\rho_c^\mathrm{ini}}}

\newcommand{\mods}[2]{\ensuremath{\mathrm{rho}#1\_\mathrm{r}#2\_\mathrm{sfs}}}
\newcommand{\modst}{\ensuremath{\mathrm{rho}9.964\_\mathrm{r}35\_\mathrm{sfs}}}
\newcommand{\modlach}{\ensuremath{\mathrm{r}10\_\mathrm{d}2.0\_\mathrm{Z}}}

\defcitealias{timmes1992a}{TW92}
\defcitealias{schwab2020a}{S20}

\usepackage{hyperref}
\hypersetup{
  colorlinks,
  citecolor=blue,
  linkcolor=blue,
  urlcolor=blue}

\begin{document} 

   \title{Observational signatures of thermonuclear electron-capture supernovae}
   \subtitle{\ion{Ne}{ii} line strengthening and color evolution as traces of the explosion mechanism} 
   \titlerunning{Observational signatures of thermonuclear electron-capture supernovae}
   \authorrunning{Holas et al.}

   \author{
        Alexander Holas\inst{1,*}
        \and
        Fionntan P. Callan\inst{2}
        \and
        Samuel W. Jones \inst{3,*}
        \and
        Friedrich K. Röpke\inst{1,4}
        \and
        Rüdiger Pakmor\inst{5}
        \and
        Alexandra Kozyreva\inst{1,8}
        \and
        Christine E. Collins\inst{6}
        \and
        Luke J. Shingles\inst{2,7}
        \and
        Stuart A. Sim\inst{2}
        \and
        Joshua M. Pollin\inst{9}
    }

   \institute{
        Heidelberger Institut für Theoretische Studien,
        Schloss-Wolfsbrunnenweg 35, 69118 Heidelberg, Germany\\
        \email{alexander.holas@mailbox.org}
        \and
        School of Mathematics and Physics, Queen's University Belfast, University   Road, Belfast BT7 1NN, UK 
        \and
        Theoretical Division, Los Alamos National Laboratory, Los Alamos, NM 87545, US
        \and
        Zentrum f\"ur Astronomie der Universit\"at Heidelberg,
        Institut f\"ur Theoretische Astrophysik, 
        Philosophenweg 12,
        69120 Heidelberg, Germany
        \and
        Max-Planck-Institut f\"ur Astrophysik,
        Karl-Schwarzschild-Str. 1, 85748 Garching,
        Germany
        \and
        School of Physics, Trinity College Dublin, The University of Dublin, Dublin 2, D02 PN40, Ireland
        \and
        GSI Helmholtzzentrum für Schwerionenforschung, Planckstraße 1, 64291, Darmstadt, Germany
        \and
        Sapienza Università di Roma, Dipartimento di Fisica, 00185, Roma, Italy
        \and
        Department of Physics, Oregon State University, 301 Weniger Hall, Corvallis, OR 97331-6507, USA\\
        $^*$The NuGrid collaboration
    }

   \date{Received September 15, 1996; accepted March 16, 1997}    

    \abstract{
    Thermonuclear electron-capture supernovae (tECSNe) are a potential fate of certain intermediate mass stars forming ONe cores at the end of their evolution. While simulations suggest that these explosions are a viable alternative to their collapsing counterpart, no synthetic observables exist that would allow for their identification among observed transients.
    }{
    In this work, we present first of their kind synthetic observables of a tECSN explosion simulation, aiming to establish whether these explosions can occur in nature. Additionally, we investigate potential observational signatures that could be used to separate these explosions from other similar astronomical transients such as pure deflagrations in CO white dwarfs.
    }{
    We carry out 3D photospheric phase and 1D late phase simulations using the Monte Carlo radiative transfer code \textsc{Artis}.
    As input, we use a tECSN explosion simulation and a CO deflagration simulation with comparable ${^{56}}$Ni production, both of which have been computed with the \textsc{Leafs} code.
    }{
    We find that both models have similar observational characteristics, akin to SNe~Iax-like events. The ejecta of the tECSN model are characterized by a $M({^{56}}\mathrm{Ni})/M_\mathrm{ej}$ ratio $25\,\%$ lower than that of comparable CO deflagration models. At early times, the tECSN model is characterized by a slower decline in the red colors compared to the CO deflagration due the greater amount of Ti and Cr synthesized in the tECSN explosion, leading to an increased absorption from these elements at blue wavelengths and subsequent fluorescence to red wavelengths. At late times, the tECSN model exhibits an exceptionally strong $12.8\,\mu$m \ion{Ne}{ii} emission line, which strengthens substantially over time, whereas its strength remains largely unchanged in the CO deflagration model.
    }{
    Our results suggest that tECSNe could potentially result in SN~Iax-like transients. Importantly, we did not find any features that are in tension with existing observables. So far, there are no indicators that unambiguously and robustly separate tECSNe from deflagrations in CO white dwarfs.
    Nonetheless, our work highlights the potential importance of the mid-infrared wavelength range for distinguishing possible explosion mechanisms.
    } 


   \maketitle
   \nolinenumbers
   
\section{Introduction}

Thermonuclear explosions of white dwarfs (WDs; \citealt{hoyle1960a}) are commonly associated with Type~Ia supernovae (SNe~Ia; \citealt{whelan1973a,iben1984a}), whereas other observed SN types are thought to be the product of stars undergoing core collapse (see, e.g., \citealt{turatto2003a,smartt2009a} for reviews).
These two explosion mechanisms (thermonuclear and core collapse) provide explanations for the explosive fate of both the low-mass ($\lesssim8\,M_\odot$) and high-mass ($\gtrsim12\,M_\odot$) end of the stellar population, respectively, although these topics remain a field of active research (see, e.g., \citealt{liu2023a}, \citealt{ruiter2025a}, and \citealt{janka2025a} for recent reviews of both explosion channels).
Left open, however, is the fate of intermediate mass stars ($8-12\,M_\odot$), as it is not clear how these stars end their lives (thermonuclear explosion or core collapse).
Originally thought to end their lives as electron capture supernovae (ECSNe; \citealt{miyaji1987a,nomoto1984a,nomoto1987a}), where electron captures at the center of the star initiate core collapse, later research (e.g., \citealt{isern1991a,canal1992a,jones2016a,jones2019a,kirsebom2019a,leung2020a}) suggested that under the right conditions a thermonuclear explosion is also a possibility.
In particular, in our recent work \citep{holas2026a}, we detail which physical mechanisms and conditions lead to either a thermonuclear ECSN (tECSN) or a collapsing ECSN (cECSN).
Current work, however, is limited to hydrodynamic simulations of the explosion, providing no predictions for the observable properties of tECSNe (apart from indirect quantities such as the ejecta composition).
\citet{jones2019b} provide some constraints on the contribution of tECSNe to galactic chemical evolution (GCE), and they find that tECSNe could occur at a low rate without introducing new tensions in the solar inventory of isotopes.
Additionally, \citet{jones2019a} find that, for example, isotopic ratios such as ${^{54}}\mathrm{Cr}/{^{52}}\mathrm{Cr}$ or ${^{50}}\mathrm{Ti}/{^{48}}\mathrm{Ti}$ present in the ejecta are in excellent agreement with isotopic ratios measured in pre-solar meteoric oxide grains.
For cECSNe, \citet{kozyreva2021a} calculated synthetic light curves that can indeed be related to existing observations.
In contrast, for tECSNe, this direct connection to observed supernovae is still missing.
The goal of this work is to establish this missing link and determine whether tECSNe can occur in nature or if there are any features that are in tension with existing observations.
Additionally, we aim to provide insights into the observational signatures of tECSNe that could be used to distinguish this explosion channel from other thermonuclear explosions.

In comparison to other astrophysical thermonuclear explosions, tECSNe are rather peculiar.
Whereas other astronomical thermonuclear explosions require an exploding WD, the progenitor evolution of tECSNe can theoretically take many shapes.\footnote{Whether or not each discussed progenitor system can reach the conditions necessary for tECSNe is still a matter of active research. For the present discussion, we simply assumed that all progenitors can somehow reach the required conditions.}
To achieve a tECSN, an ONe WD or stellar core needs reach a central density high enough for electron captures to set in. 
This usually occurs when the mass of the WD or stellar core approaches its limit for stability, the Chandrasekhar mass limit ($\mch$).
At this point, the electron capture chain ${^{20}}\mathrm{Ne}\rightarrow{^{20}}\mathrm{F}\rightarrow{^{20}}\mathrm{O}$ begins to neutronize the core, thereby further destabilizing the core and initializing collapse.
At the same time, the decay of ${^{20}}$O can ignite thermonuclear burning.
If the conditions are just right (see \citealt{holas2026a}), the thermonuclear burning can overcome the gravitational collapse and lead to a tECSN.

In general, there are three channels by which an ONe core can reach the conditions required for the initiation of an ECSN (for a review see, e.g., \citealt{doherty2017a,wang2026a}).
The first channel is by accretion-induced collapse (AIC; \citealt{nomoto1991a}) where an ONe grows close to $\mch$ by stably accreting mass from a binary companion \citep{schwab2015a}.
Another channel is a single super asymptotic giant branch (sAGB) star whose core grows close to $\mch$ by undergoing repeated unstable He shell burning cycles \citep{ritossa1999a,jones2013a}.
The third and last channel consists of an sAGB star whose envelope has been stripped through binary interaction.
Here, the resulting ONe core grows close to $\mch$ through mass accretion from a close binary companion, leading to stable He shell burning \citep{podsiadlowski2004a,tauris2015a}.
\citet{jones2019a} find that the binary sAGB star channel is by far the most likely scenario, followed by AIC, with a single star ECSN progenitor being the least likely channel.

The path by which the progenitor reaches the ignition density (e.g., single star, binary, or accretion) has limited impact on the explosion simulation,\footnote{Different progenitors can impart a different internal composition. However, to our knowledge this effect is subdominant to the ignition density itself; see \citet{holas2026a}.} but can have a drastic effect on the resulting synthetic observables.
For example, the H envelope in the single sAGB star scenario will create a plateau-like light curve due to H recombination \citep{kozyreva2024a} that is akin to a \mbox{Type~II-P} supernova (SN~II-P).
Similarly, in the binary sAGB star scenario, the He shell burning can leave behind substantial amounts of C on the surface that can contaminate the resulting synthetic observables, as tECSNe produce little C during the explosion themselves.
Since the progenitor scenario is quite uncertain, we mostly focus on the explosion of the bare ONe core, highlighting signatures of the tECSN explosion mechanism itself rather than the impact of various progenitor scenarios.

Historically, pure deflagrations in $\mch$ CO WDs have been linked to SNe~Iax \citep{branch2004a,jordan2012a,kromer2013a,fink2014a,lach2022a}, also known as 02cx-like SNe~Ia \citep{li2003a,foley2013a}, an SN~Ia subtype characterized by a lower luminosity and kinetic ejecta energy compared to normal SNe~Ia (see, e.g., \citealt{jha2017a} for a review).
Pure deflagrations in CO WDs reproduce several observed features of SNe~Iax, particularly their low ejecta velocities.
However, they struggle to reproduce others such as the R band decline rate \citep{lach2022a,callan2024a}.
\citet{callan2024a} demonstrated that radiative transfer (RT) simulations including the contribution of a luminous remnant predicted by CO deflagration explosion models improved the agreement with observed SNe~Iax, particularly of the decline rate in the redder bands.
The nature and evolution of this remnant, however, is still very uncertain, and many open questions remain, but we do not consider the impact of the bound remnant further in this work.

In Sect.~\ref{sec:methods} we outline the methods used for the simulations discussed in this work.
In Sect.~\ref{sec:nucleo} we discuss the nucleosynthetic composition and structure of the explosion models presented in this work.
Early-time observables are discussed in Sect.~\ref{sec:early}, while late-time observables are compared in Sect.~\ref{sec:late}.
Lastly, we briefly discuss the potential impact of a H shell on the observed light curve in Sect.~\ref{sec:progenitor} before concluding in Sect.~\ref{sec:conclusion}.

\section{Methods}~\label{sec:methods}

In this work we focus on a tECSN simulation using the same setup as described by \citet{holas2026a}; see Sect.~\ref{sec:expl_sim}.
For comparison, we also calculated synthetic observables for the \modlach{} model (also referred to as ``the CO deflagration model'') of \citet{lach2022b}.\footnote{Preliminary experiments showed that other CO deflagration models of comparable luminosity such as the N1def model of \citet{fink2014a} yield similar results. Due to limited computational resources we restricted our comparison to the \modlach{} model, but we expect that our conclusions transfer to other pure deflagration models as well.}
This model describes a pure deflagration in a $\mch$ WD with a $50\%/50\%$ (by mass) CO composition, $\rhoini = 2.0\times10^9\,\mathrm{g}\,\mathrm{cm}^{-3}$, ignited $10\,\mathrm{km}$ off center.
For more detail on this model, we refer the reader to \citet{lach2022b}.
It was chosen for comparison because it produces similar amounts of $0.06\,\mathrm{M_\odot}$ of ${^{56}}$Ni and both the nuclear and kinetic energy released during the explosion is comparable to that of our tECSN model (see Table~\ref{tab:radioactives}).
Here, we recalculated both the nucleosynthesis and RT using the methods described in the following sections, based on the original hydrodynamic explosion simulation described by \citet{lach2022b}.
This is to minimize the impact of differences in, for example, the size of the network or the atomic data set on the resulting synthetic observables.
Particularly the size of the atomic data set can have a noticeable impact on quantities such as the decline rate; see \citet[Fig. B.2]{holas2025a}.
Note that for our re-processed version of the \modlach{} model, we initially ignored the impact of solar metallicity (see Sect.~\ref{sec:nucleosynthesis}), which is included in the post-processing of \citet{lach2022b}, again in an attempt to isolate the effect of the explosion mechanism.

In addition to the tECSN and CO deflagration model, we included several variations of these models to gauge the potential impact of C and Ne introduced by the progenitor evolution rather than the explosion itself.
This allowed us to evaluate whether or not certain features caused by the explosion mechanism could also be produced by this progenitor contamination.
To estimate the impact of these elements, we reprocessed the tracer particles (see Sect.~\ref{sec:nucleosynthesis}), where we set the initial abundances according to the studied case, that is either with initial C or Ne.
This way we are able to capture the effect of nuclear burning on these elements without re-running the explosion simulation directly.
We deem this a sufficient approximation as these trace abundances will not significantly impact the energy generation or the flame propagation.

\subsection{Explosion simulation}~\label{sec:expl_sim}

The tECSN simulation studied in this work was simulated using the \textsc{Leafs} code as described by \citet{holas2026a}.
Here, we only briefly summarize the details of the setup and refer the reader to \citet{holas2026a} for more details.
\textsc{Leafs} is based on the \textsc{Prometheus} code \citep{fryxell1989a} and it solves the reactive Euler equation in 3D using a piecewise parabolic scheme \citep{colella1984a}.
The code tracks both the flame front and the expansion of the ejecta by nesting an expanding uniform Cartesian grid inside a non-uniform Cartesian grid \citep{ropke2006a}.
A level-set \citep{osher1988a} is used to approximate the flame front \citep{reinecke1999b}, where the laminar flame speed follows the parameterization of \citet{schwab2020a}.
The contribution of unresolved velocity fluctuations to the effective flame speed is modeled through a turbulent subgrid scale model \citep{schmidt2006a,schmidt2006b}.
\textsc{Leafs} implements a Helmholtz equation of state (EOS) described by \citet{timmes2000a}, including Coulomb corrections \citep{yakovlev1989a}.
Electron captures are accounted for following the prescription of \citet{jones2016a}.
For more details on the numerical setup, we refer the reader to \citet[Sect. 2.2]{holas2026a}.

The initial conditions for the model in this work were based on \citet{kirsebom2019a}.
Specifically, we approximated their model that predicts an off-center ignition at $35\,\mathrm{km}$.
Including their new measurement of the forbidden transition between the ground states of ${^{20}}\mathrm{Ne}$ and ${^{20}}\mathrm{F}$, this ignition occurs at a central density of $\log\rho_c^\mathrm{ini}\approx9.964\,\mathrm{g}\,\mathrm{cm}{^{-3}}$.
Using these parameters, we set up an initial $65\%/35\%$ ONe WD as described in \citet{holas2026a} and artificially ignite it $35\,\mathrm{km}$ off-center.
Following the nomenclature of \citet{holas2026a}, we call this model the \mods{9.964}{35} model (or alternatively ``the tECSN model'').

\subsection{Nucleosynthesis}~\label{sec:nucleosynthesis}
To obtain a detailed nucleosynthetic composition of the ejecta, we performed nucleosynthetic post-processing.
We followed the approach established by other works (see, e.g., \citealt{seitenzahl2013a,lach2022b}) of distributing around $10^{6}$ variable-mass Lagrangian tracer particles that get advected in the flow as the WD explodes.
Here, up to a mass coordinate of $0.8\,\msun$, tracer particles were distributed equal by mass, followed by particles distributed equal by volume up to $1.3\,\msun$, and the remaining tracer again equal by mass.\footnote{We note that the \modlach{} model of \citet{lach2022a} uses $4\times10^6$ equal by mass tracer particles instead.}
These tracers record the thermodynamic conditions along their trajectories, which in turn are fed into a nuclear post-processing network.
Here, we used the NuGrid\footnote{\url{nugridstars.org}} nuclear reaction network \citep{pignatari2016a,ritter2018a,jones2019c} with modifications described by \citet{jones2019a}.
The network includes a total of $5213$ species as specified in \citet[table 2]{jones2019a}.
Due to the extreme conditions encountered in ECSNe, a large network is required to adequately cover the nuclear statistical equilibrium (NSE) solutions for electron fractions in the range of $0.25 \leq Y_e \leq 0.5$ and to avoid issues at the network boundaries.
Particular care has been taken in selecting weak reaction rates, that is, electron and positron captures as well as $\beta^\pm$-decays.
The weak reaction rate composite table is illustrated in \citet[Fig. 3]{jones2019a}; it is made up of rates from \citet{caughlan1988a}, \citet{takahashi1987a,goriely1999a}, \citet{oda1994a}, \citet{fuller1985a}, \citet{langanke2000a}, \citet{nabi2004a}, and \citet{cyburt2010a}.
For more details on the nuclear reaction network, we refer the reader to \citet{jones2019a} and references therein.

To study the impact of residual C, we computed two model variations of the nucleosynthetic post-processing, \modst\_0.01C and \modst\_0.01C\_0.5Cshell, based on the \modst{} explosion simulation.
For these models, we used the calculations of \citet{schwab2019b} as a reference, where they investigate the impact of residual C on the evolution of ONe cores.
Here, they place an upper limit of $X({^{12}}\mathrm{C})=0.01$ in the core.
For abundances higher than that, C will be ignited before reaching densities typical of the tECSN regime studied here (see also, e.g., \citealt{antoniadis2020a}).
Therefore, we used this value as an upper boundary for the core contamination.
In the case of the C shell, \citet{schwab2019b} do not give exact values, but their Figs.~1 and 3 suggest $X({^{12}}\mathrm{C})\approx 0.5$ for mass coordinates $M(r) > 1.2\,\msun$.
We again adopted this value as a rough estimate of the shell composition.
From these values, we calculated two models.
The first model, \modst\_0.01C, only includes the core contamination, whereas the second model, \modst\_0.01C\_0.5Cshell, includes both the core and shell contribution.
In both cases, we have replaced O and Ne in equal parts by mass.

Similar to the investigation of residual C, we computed model variations of the \modlach{} model, modifying the initial ${^{22}}\mathrm{Ne}$ content, the \modlach{}\_Ne\_1.0Z and the \modlach{}\_Ne\_5.0Z model.
Here, we enhanced the initially zero-metallicity abundance by $1\times Z_\odot$ and $5\times Z_\odot$ worth of ${^{22}}$Ne, with $Z_\odot \approx 0.02$ \citep{asplund2009a}, reducing the C and O mass fraction accordingly by equal parts.
We note that while in the \modlach{}\_Ne\_1.0Z case, the addition of ${^{22}}$Ne should have a negligible effect on the explosion simulation, in the \modlach{}\_Ne\_5.0Z case this addition will likely have a noticeable effect on the overall explosion (predominately by a reduced energy release) which would result in different thermodynamic trajectories compared to the original \modlach{} model.
Therefore, the results obtained from this model are affected by comparably large uncertainties.

Following the nucleosynthetic post-processing step, we mapped the abundances to a Cartesian grid required for the subsequent RT calculations.
In this step, we used the density field of the hydrodynamic simulation and the abundance pattern of the post-processed tracer particles.
Here, we removed the bound\footnote{We define bound material as material with a negative total energy, $E_\mathrm{kin}+E_\mathrm{int}+E_\mathrm{pot} < 0$.} remnant in both the \mods{9.964}{35} and \modlach{} case.
We filled the void using nearest neighbor interpolation for the density field, while the abundances were mapped from the remaining unbound tracer particles, effectively clipping the density in this region to its boundary value.
This changes the total ejecta mass and adds additional radioactive material such as ${^{56}}$Ni.
However, we found that this does not significantly impact our results compared to the case where we fill the bound remnant region with a vacuum.
The input model model contains elements up to Zr, with heavier elements (which have a negligible mass fraction) being added to Fe instead.
For early time observables, we mapped to a 3D Cartesian grid with $50^3$ cells, whereas for late-time observables, we used a 1D model with $100$ radially averaged shells.

\subsection{Radiative transfer}
\subsubsection{\textsc{Artis}}
We obtained synthetic observables using the 3D, time-dependent Monte Carlo RT code \textsc{Artis} \citep{sim2007a,kromer2009a,bulla2015a,shingles2020a,artiscollaboration2025a}, based on the methods described by \citet{lucy2002a,lucy2003a,lucy2005a}.
In this work, we focus both on early time photospheric observables, as well as late-time nebular observables.
For photospheric observables, we used the approximate non local thermal equilibrium (LTE) prescription outlined by \citet{kromer2009a}, where the ionization state is estimated based on photoionization rates evaluated from the simulation.
Here, optically thick cells are treated using a gray approximation.
For more details on this version and caveats of the approximate non-LTE treatment (particularly in the near-infrared; NIR) we refer the reader to \citet{kromer2009a} and \citet[Sect. 2.3]{holas2025a}.
In the case of the photospheric phase, we obtained synthetic observables using $10^8$ energy packets and $100$ logarithmically spaced time steps between $2$ and $60$ days after the explosion, assuming continuous homologous expansion.
For the photospheric phase, we utilized the \textsc{big\_gf-4} atomic data set described by \citet{kromer2009a} which is based on the atomic data sets of \citet{kurucz1995a,kurucz2006a} and contains $142$ ions, $8.3\times10^6$ atomic lines with $1.3\times10^5$ levels and $3.6\times10^4$ photoionization transitions.

The code configuration used for nebular phase quantities uses the full non thermal and non-LTE capabilities of \textsc{Artis} and described by \citet{shingles2020a}.
This includes a full non-LTE population and ionization solver.
Here, we accounted for up to $100$ non-LTE levels for elements with $Z<20$, and $200$ levels for heavier elements.
Additionally, the changes implemented by \citet{shingles2020a} include routines to deal with collisions with non thermal leptons.
Importantly, \textsc{Artis} solves the Spencer-Fano equation \citep{kozma1992a} in order to calculate the energy distribution of high-energy leptons.
For more details, see, for example, \citet{shingles2020a}, \citet{callan2025a}, and \citet{pollin2025a}.
Here, we used $3\times10^9$ energy packets and $150$ logarithmically spaced timesteps from $50$ to $200$ days after the explosion.
As for the photospheric phase, we assumed homologous expansion for the entire duration.
For the nebular phase, we instead used an atomic data set based on the compilation of \textsc{CMFGEN} \citep{hillier1990a,hillier1998a,blondin2023a}.
This data contains $4.1\times10^6$ lines with $4.1\times10^4$ level, $1.1\times10^5$ photoionization transitions, and $70$ ions.
For more details, see \citet{callan2025a}.

In both cases \textsc{Artis} tracks the decay of all radioactive isotopes \citep{shingles2023a} obtained from the nucleosynthetic post-processing step up to ${^{107}}$Zr. 
We truncated the atomic data in both the early- and late-time cases at Zn due to the lack of sufficient data for heavier elements, for example photoionization cross sections required for the approximate non-LTE treatment.
Tests with larger atomic data sets in pure LTE showed that elements heavier than Fe have little to no impact on the resulting observables and we therefore deem this a sufficient approximation.
While we were unable to establish the impact of elements such as Sr on the late-time spectra (due to unavailability of the necessary atomic data), given the low mass fraction we do not expect these heavier elements to play a significant role in the spectral appearance.

As mentioned at the end of Sect.~\ref{sec:nucleosynthesis}, the early phase simulations were conducted in 3D and the late phase simulations in 1D.
At early times we can utilize the approximate non-LTE treatment for a 3D simulation. 
At late phases this approximate non-LTE treatment is no longer sufficient due to such issues as a lack of collisions, non thermal effects, which are important in the nebular phase.
Therefore we needed to use the full non-LTE treatment.
Full non-LTE in 3D is computationally expensive and outside the scope of this study, and therefore we limited ourselves to determining the general characteristics and signatures of the 1D models in the nebular phase.
However, we expect that full 3D calculations will be needed to eventually fully reconcile observations with models.
For example, observed SN~Iax nebular profiles exhibit pronounced effects of an asymmetric explosion \citep{kwok2025a}.

\subsubsection{\textsc{Stella}}
In Sect.~\ref{sec:progenitor} we also briefly illustrate the impact of a potential H envelope on the resulting light curves.
Here, we closely followed the approach described by \citet{kozyreva2024a} and refer the reader to their work for more details on the methods.
In short, we utilized a spherically averaged ejecta model (the same as is used for the non-LTE \textsc{Artis} calculations) of the bare core explosion inside the $8.75\,M_{\odot}$ stellar envelope model of \citet{jones2013a}.
This stellar model has been calculated using \textsc{mesa} \citep{paxton2011a,paxton2013a,paxton2015a,paxton2018a,paxton2019a} and has been used as reference when constructing the initial WD used in our tECSN explosion models.
The combined explosion and envelope model is then used as input for 1D RT calculations using \textsc{Stella}\footnote{The version used in this work is not the same as can be found in \textsc{mesa}, but a private version.} \citep{blinnikov1993a,blinnikov2006a}.
Importantly, \textsc{Stella} models the hydrodynamic shock propagation of the ejecta with the surrounding envelope and the resulting impact on the light curves.

\section{Nucleosynthesis and ejecta structure}\label{sec:nucleo}

\subsection{Ejecta structure}
\begin{figure*}
    \centering
    \includegraphics[]{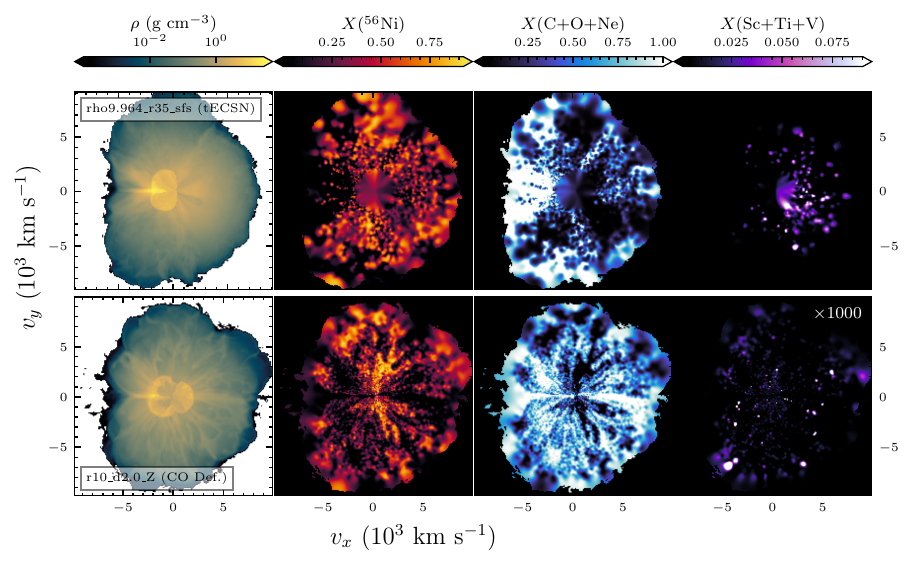}
    \caption{Slices of the mapped ejecta structure of both explosion models at $100\,\mathrm{s}$ after the explosion. Here, we show the density, ${^{56}}$Ni mass fraction, the unburnt material mass fraction (represented by C, O, and Ne), and the combined Sc, Ti, and V mass fraction. Note that for the \modlach{} model, we have enhanced the mass fraction of Sc, Ti, and V by a factor $1000$ for better visibility. The depicted slices are at a resolution of $200^3$; the actual input for model for \textsc{Artis} has a resolution of $50^3$ in 3D and $100$ radially averaged shells in 1D.}
    \label{fig:model}
\end{figure*}
Before examining the resulting synthetic observables, we briefly compare the nucleosynthetic yields and ejecta structure of the \modst{} and \modlach{} models.
Fig.~\ref{fig:model} illustrates the mapped ejecta for both models.
Here, it can immediately be seen that both explosion models exhibit overall a rather similar structure.
However, there are a few critical differences.
First, the tECSN model is somewhat more compact compared to the CO WD deflagration model (note the different box sizes in Fig.~\ref{fig:model}), resulting in different density profiles.
In the leftmost density column, the impact of the bound remnant can be clearly seen in the central high-density region.
This is caused by the ejection of initially bound material that fell back onto the central object and got dynamically ejected during the formation of the bound central remnant.
These ejected outer layers are formally unbound and are used to fill the bound region during our mapping routine.

Second, we note that the ejecta are in both cases reasonably well mixed, as is generally expected from pure deflagrations \citep[e.g.,][]{travaglio2004a,ropke2007a,fink2014a,lach2022b}.
In Fig.~\ref{fig:model}, in the two middle columns, we illustrate both the ash composition, represented by the ${^{56}}$Ni mass fraction, and the fuel contribution, represented by the C, O, and Ne mass fraction,\footnote{In principle, Ne is not part of the fuel in the \modlach{} model and the \modst{} model has no initial C. However, for illustration purposes, we add them together. For the present qualitative discussion this is sufficient due to the respectively low C and Ne production during the explosion. See Fig.~\ref{fig:artis_c_ne_app}.} showing that both fuel and ashes are well mixed.
The largest anomaly is the negative x-direction of the \modst{} model, where little ash and substantial amounts of fuel can be found.
This is caused by the asymmetric ignition, leading to the flame rising in the positive x-direction, eventually burning around the core region.
Here, the less energetic ONe deflagration does not fully burn around the core, leaving some unburnt material, whereas the CO deflagration fully engulfed the core region.
We note that this that this column of unburnt material is also observed in other pure deflagration models, for example the N20def model of \citet{fink2014a}.
For a discussion of its impact on spectral features, see, for example, \citet{kwok2025a}.

Third, we highlight the production and structure of elements such a Sc, Ti, and V, shown in the rightmost column of Fig.~\ref{fig:model}.
Apart from the obvious difference in overall production (note that the mass fractions for the \modlach{} model were scaled by a factor of $1000$ for better visibility), it also stands out that these elements can be found predominantly in the high to medium density regions of the tECSNe models, with the low-density regions showing little production.
In contrast, the CO deflagration model shows a more homogeneous distribution of these elements, although this is likely more connected to the fact that these elements a scarcely produced in this explosion channel.
We note that the high concentration of these elements in the center of the \modst{} model is partially caused by the nearest neighbor interpolation of the bound remnant region.
However, we find that removing this material makes little difference on the trends discussed in the following sections.
These elements are the first key difference between CO deflagrations and high-density ONe deflagrations, i.e., tECSNe, that will have a noticeable effect on the observable quantities, as we will see in the later Sect.s~\ref{sec:early} and \ref{sec:late}.

Lastly, we also briefly mention the ejecta structure of the model variants, illustrated in Fig.~\ref{fig:artis_c_ne_app}, the C and Ne distribution specifically.
In case of C, the effective absence of C in the original \modst{} model clearly stands out.
In both \modst{} variations, \modst{}\_0.01C and \modst{}\_0.01C\_0.5Cshell, the respective initial C abundance can be identified in the outer ejecta layers.
However, the central regions exhibit a comparably low C abundance, suggesting that most C in the core is burned durning the explosion and only C in the shell gets mixed into the ejecta.
This is further substantiated by the fact that in the \modst\_0.01C and \modst\_0.01C\_0.5Cshell the remaining C is mostly co-spacial, implying that this C indeed originated from the shell region.
Regarding the Ne variations of the \modlach{} model, we observe a similar behavior, that is the Ne mostly survives in the unburnt ashes.

\subsection{Nucleosynthesis}\label{sec:yields}
\begin{figure}
    \centering
    \includegraphics[]{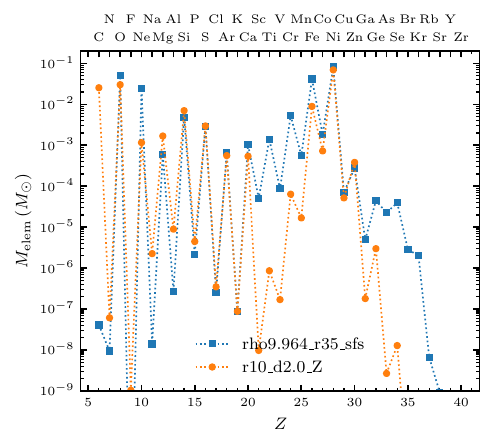}
    \caption{Elemental masses contained in the ejecta of both masses contained in the \textsc{Artis} model at $100\,\mathrm{s}$ after the explosion.}
    \label{fig:elem_abu}
\end{figure}
Fig.~\ref{fig:elem_abu} illustrates the elemental masses of both models.
Similar to the ejecta structure, both models are, broadly speaking, similar, particularly for lighter elements up to Ca.
In this region, the largest differences can be found in C and Ne, which can be attributed to the initial composition; as both explosion models burn a considerable fraction of the ejected mass to NSE, little of these elements are produced during the explosion.
Apart from these two elements, there are noticeable differences in the production of elements such as Mg (note the logarithmic scale of Fig.~\ref{fig:elem_abu}) and Al.
However, as we will see later in Sect.~\ref{sec:early}, these differences are sub-dominant for the RT calculations compared to the strong difference found in the Sc to Mn range.
Here, the CO deflagration model effectively produces none of these elements, particularly Ti, which will be relevant for the color evolution (see Sect.~\ref{sec:early}) as it can cause, for example, substantial line blanketing at blue wavelengths.
Finally, for elements with $Z>30$ in the Zn to Zr range, we find that the tECSN model produces substantially more of these elements as is characteristic for tECSNe \citep{jones2019a}.

As far as our model variations are concerned, the addition of C does not significantly impact the resulting nucleosynthesis and only the composition of the unburned ejecta changes (most notably in the \modst{}\_0.01C\_0.5Cshell model).
In contrast, the addition of ${^{22}}$Ne has a much more pronounced impact on the subsequent nucleosynthesis.
In particular, it reduces the overall production of ${^{56}}$Ni quite significantly.
The total ${^{56}}$Ni yield only amounts to $5.74\times 10^{-2}\,M_\odot$ and $4.53\times 10^{-2}\,M_\odot$ for the \modlach{}\_Ne\_1.0Z and \modlach{}\_Ne\_5.0Z variant, respectively.
Particularly in the latter, this implies that the introduction of ${^{22}}$Ne in the initial composition will most likely also impact the explosion as a whole, an effect that we cannot capture by only rerunning the nucleosynthetic post-processing stage.

\begin{table}
\caption{Mass (in $M_\odot$) of select radioactive isotopes contained in both explosion models as well as their half-lives and various ejecta properties.}
\label{tab:radioactives}
\centering
\setlength{\tabcolsep}{4pt}
\begin{tabular}{lccc}
\hline\hline
Isotope & \mods{9.964}{35} & \modlach{} & $t_{1/2}$ (days)\\
\hline
${^{56}}\mathrm{Ni}$ & 6.29e-02 & 5.90e-02 & 6.08 \\
${^{57}}\mathrm{Ni}$ & 1.95e-03 & 1.82e-03 & 1.48 \\
${^{55}}\mathrm{Co}$ & 1.37e-03 & 6.24e-04 & 0.73 \\
${^{66}}\mathrm{Ni}$ & 1.31e-03 & 8.96e-14 & 2.27 \\
${^{52}}\mathrm{Fe}$ & 6.54e-04 & 3.76e-04 & 0.34 \\
${^{49}}\mathrm{Sc}$ & 5.04e-05 & 3.00e-12 & 0.04 \\
${^{62}}\mathrm{Zn}$ & 2.95e-05 & 4.47e-05 & 0.38 \\
${^{56}}\mathrm{Co}$ & 2.95e-05 & 1.20e-05 & 77.23 \\
${^{61}}\mathrm{Co}$ & 2.54e-05 & 5.47e-09 & 0.07 \\
${^{59}}\mathrm{Fe}$ & 1.97e-05 & 1.12e-09 & 44.49 \\
${^{48}}\mathrm{Cr}$ & 1.74e-05 & 1.44e-05 & 0.90 \\
${^{78}}\mathrm{Ge}$ & 1.23e-05 & 1.36e-25 & 0.06 \\
${^{72}}\mathrm{Zn}$ & 6.70e-06 & 3.49e-21 & 1.94 \\
${^{65}}\mathrm{Ni}$ & 6.05e-06 & 1.27e-13 & 0.10 \\
${^{56}}\mathrm{Mn}$ & 5.97e-06 & 2.93e-09 & 0.11 \\
${^{67}}\mathrm{Cu}$ & 5.62e-06 & 4.65e-13 & 2.58 \\
${^{58}}\mathrm{Co}$ & 2.51e-06 & 3.62e-07 & 70.86 \\
${^{51}}\mathrm{Cr}$ & 1.92e-06 & 1.93e-07 & 27.70 \\
${^{61}}\mathrm{Cu}$ & 9.62e-07 & 1.73e-06 & 0.14 \\
${^{52}}\mathrm{Mn}$ & 7.09e-07 & 3.20e-07 & 5.59 \\
\hline
$M_{\mathrm{rad}}$ ($\msun$) & 6.84e-02 & 6.19e-02 & ... \\
\hline
$M_{\mathrm{ej}}$ ($\msun$) & 2.17e-01 & 1.49e-01 & ... \\
\hline
$E_\mathrm{kin}$ ($10^{50}\,\mathrm{erg}$) & 0.28 & 0.29 & ... \\
$E_\mathrm{nuc}$ ($10^{50}\,\mathrm{erg}$) & 4.61 & 2.32 & ... \\
\hline
\end{tabular}
\end{table}
Lastly, we briefly discuss the radioactive elements found in both deflagration models.
Table~\ref{tab:radioactives} lists the ejected mass of the most abundant radioactives of both models with a half life of fewer than $100$ days and more than an hour (with the exception of ${^{49}}$Sc), taken at $100\,\mathrm{s}$ after the explosion.
As already mentioned in the introductory paragraph of Sect.~\ref{sec:methods}, here it can be seen that both models produce roughly similar amounts of ${^{56}}$Ni, approximately $0.06\,M_\odot$ and therefore one would expect a similar observable brightness for both events.
In case of the CO deflagration model, the radioactive ejecta mass is dominated by ${^{56}}$Ni, with a secondary contribution from ${^{57}}$Ni, ${^{52}}$Fe, and ${^{55}}$Co.

In contrast, the ejecta of the \modst{} model has many more radioactive elements that contribute noticeably to the total radioactive mass, in particular elements such as ${^{66}}$Ni and ${^{59}}$Fe.
This makes up around a $4\,\%$ difference in the total radioactive mass (not counting the around $6\,\%$ difference in the ${^{56}}$Ni mass).
However, most isotopes have rather short half lives and will decay within the first few days, while longer lived isotopes (for example ${^{59}}$Fe) are sub-dominant with respect to the ${^{56}}$Ni mass.
Interestingly, the radioactive material makes up around $42\,\%$ of the total ejecta mass in the \modlach{} case, whereas it only constitutes $32\,\%$ in the \modst{} case, despite the rather similar ${^{56}}$Ni mass.
This is appears to be a rather significant difference between these explosion channels.
In particular, \citet{lach2022b} find that their models produce effectively the same $M({^{56}}\mathrm{Ni})/M_\mathrm{ej}$ ratios of around $0.4$, which appears to be in tension with several (particularly fainter) observed transients; see their Figs.~4 and 5.
Therefore, tECSNe could provide an possible alternative explosion mechanism for objects with a low $M({^{56}}\mathrm{Ni})/M_\mathrm{ej}$ ratio and could potentially be identified by this diagnostic value.
For the models computed by \citet{holas2026a}, we find an average $M({^{56}}\mathrm{Ni})/M_\mathrm{ej}$ ratio of around $0.25-0.3$ (Holas et al., in prep.).

Although from the perspective of energy deposition both cases are dominated by the ${^{56}}$Ni decay chain, it is important to properly track all radioactive decays (beyond the initially post-processed $100\,\mathrm{s}$ after the explosion) for the tECSN case as the elemental composition will slightly change over time.
In previous works concerning SN Ia simulations \citep[e.g.,][]{sim2007a,pakmor2010a,kromer2015a,holas2025a}, only decays of ${^{56}}$Ni, ${^{56}}$Co, ${^{52}}$Fe, and ${^{48}}$Cr were considered during the RT calculations, which for those scenarios was clearly sufficient given that other isotopes at most contribute at a $1\,\%$ level.
This is the case for the \modlach{} model as well.
However, for our tECSN model the error would be much more noticeable at the $10\,\%$ level, particularly through species such as ${^{66}}$Ni and therefore we tracked the decay of all relevant radioactive isotopes in the present work.

In this context we particularly highlight ${^{49}}$Sc.
Although this isotope is rather short lived, it needs to be properly decayed for two reasons.
On the one hand, if not properly decayed, it will result in noticeable Sc features in the resulting spectra of the \modst{} model, as it constitutes the majority of all Sc in the ejecta.
On the other hand, ${^{49}}$Sc decays to ${^{49}}$Ti, thereby further enhancing the Ti abundance in the tECSN model.

In summary, the ejecta of the \modst{} and \modlach{} models are fairly similar in their structure and composition.
However, the tECSN model is somewhat more compact and contains significantly more material in the Sc to Mn range.
Moreover, the isotopic composition of the \modst{} model is much more complex, particularly regarding the radioactive isotopes, which requires more care in accurately representing the composition at the various phases during the RT calculation.

\section{Early-time observables}\label{sec:early}
In this section we examine the early-time synthetic observables of the \modst{} and \modlach{} models.
We focus on a comparison between the two models and signatures of the explosion mechanism.
In several places, we also compare our synthetic observables to observed SNe.
However, in these cases we focus on how either explosion mechanism can explain certain features in those observations, rather than attempting to match our models to individual objects.

\subsection{Light curves}\label{sec:early_lc}
\begin{figure*}
    \centering
    \includegraphics[]{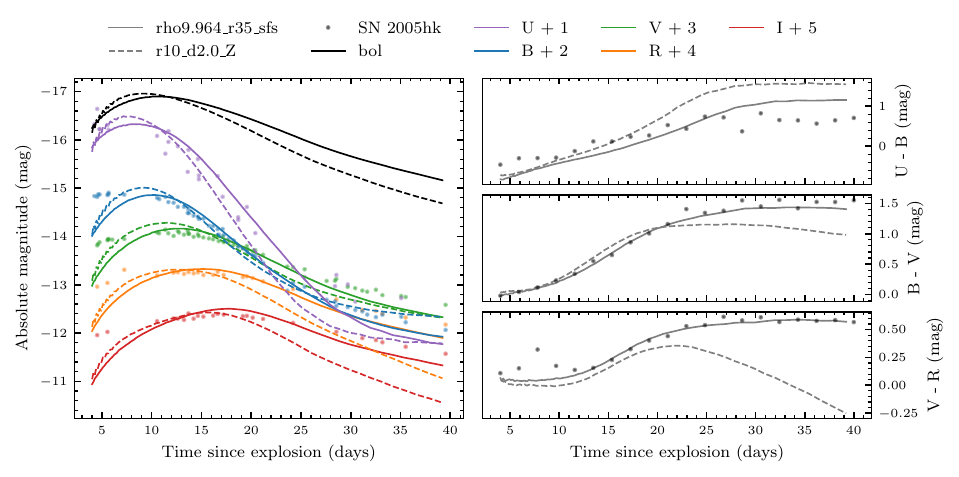}
    \caption{Synthetic early-time light curves of both explosion along selected colors. For comparison, we also show observational data of SN~2005hk (taken from the Open Supernova Catalog; \citealt{guillochon2017a}). For ease of comparison, we have adjusted the magnitude (the luminosity distance to be precise) and phase of SN~2005hk to better overlap with our synthetic light curves. The individual light curves have been shifted for better visibility. For the associated viewing angle dispersion, see Fig.~\ref{fig:lightcurves_va}.}
    \label{fig:lightcurves}
\end{figure*}
\begin{table}
\caption{Key early-time angle-averaged light curve properties of both explosion models.}
\label{tab:lc_props}
\centering
\begin{tabular}{lcc}
\hline\hline
 & \mods{9.964}{35} & \modlach{} \\
\hline
$M_{\mathrm{bol, max}}$ (mag) & -16.90 & -16.96 \\
$M_{\mathrm{U, max}}$ (mag) & -17.33 & -17.49 \\
$M_{\mathrm{B, max}}$ (mag) & -16.86 & -17.01 \\
$M_{\mathrm{V, max}}$ (mag) & -17.16 & -17.28 \\
$M_{\mathrm{R, max}}$ (mag) & -17.33 & -17.31 \\
$M_{\mathrm{I, max}}$ (mag) & -17.50 & -17.43 \\
$\Delta m_{15}(B)$ (mag) & 1.99 & 2.08 \\
$t_\mathrm{rise}(\mathrm{bol})$ (days) & 11.75 & 10.38 \\
\hline
\end{tabular}
\end{table}
In Fig.~\ref{fig:lightcurves}, we illustrate the light curves of the two models as well as an observed light curve of the prototypical bright SN~Iax SN~2005hk \citep{hicken2009a,silverman2012a,stahl2019a} for reference.
In Table~\ref{tab:lc_props}, we list some characteristic properties of the angle-averaged light curves of the two models, such as peak magnitudes, decline rates, and rise times.
As is expected from the results in the previous Sect.~\ref{sec:nucleo}, both models produce roughly the same magnitude, particularly in bolometric light.
Interestingly, the \modlach{} model is slightly brighter even though it has less radioactive material powering its light curve.
This is caused in part by the compositional differences, namely that the tECSN model has significantly more absorption from species such as Ti at bluer wavelengths which redden the spectrum and slow down the escape of photons.
Furthermore, the fraction of radioactive material in the \modlach{} model is noticeably larger than in the \modst{} model (see Sect.~\ref{sec:nucleo}), which in previous work has been shown to be a main driver behind light curve timescales \citep[e.g.,][]{lach2022b}.
Furthermore, the CO deflagration model has a less compact ejecta structure, leading to lower densities, which makes it easier for photons to escape, and hence make the explosion brighter.
These effects are also responsible for the faster rise time in case of the CO deflagration model, which reaches its bolometric peak almost $1.5\,\mathrm{days}$ faster than the tECSN model.
In contrast, the \modst{} model declines slower, particularly in the red bands.
The slower decline, as we discuss in Sect.~\ref{sec:early_spec}, is due to line blanketing in the blue and additional emission at longer wavelengths from Ti and Cr.

This brings us to the most noticeable difference between the two explosion models, their color, illustrated in the right column of Fig.~\ref{fig:lightcurves}.
At early epochs, up to around $20\,\mathrm{days}$ after the explosion, both models exhibit similar colors.
However, after this point their evolution diverges.
In particular, the \modlach{} becomes bluer in the B-V and V-R colors, especially the latter, whereas the \modst{} model remains consistently red.
The extreme change in the V-R color of the CO deflagration model is due to a loss of flux in the $5500$\,\AA\ to $8000$\,\AA\ wavelength range that is maintained by increased absorption from Ti and Cr at blue wavelengths and subsequent fluorescence of this radiation at redder wavelengths (see Sect.~\ref{sec:early_spec} and Fig.\ref{fig:spec_time_early}\footnote{We also refer the reader to \citet[Fig. 16]{lach2022b} where this relatively weak R band flux can also be observed for a similar model.}); exactly the wavelength range of the R band.
In the B-V color, the \modst{} model is slightly redder than the \modlach{} model due to additional flux from predominantly Cr fluoresence in the V band.
Finally, in the U-B color, the CO deflagration produces marginally redder colors than the tECSN, caused by Ti absorption in the tECSN model.

We also briefly compare our model light curves to observations of SN~2005hk.\footnote{We note that our results do not depend on the choice of observed SN. As far as we can tell, the color evolution exhibited by SN~2005hk is shared by a multitude of other SNe~Iax with varying peak luminosities. See, e.g., SN~2002cx \citep[Fig.~4]{li2003a}, SN~2014ck \citep[Fig.~5]{tomasella2016a}, or SN~2019muj \citep{barna2021a}.}
The models of \citet{lach2022b}, have previously been linked to SNe~Iax (of which SN~2005hk is a prototypical object) and we briefly revisit this comparison here.
For this purpose, we have scaled the peak magnitude and phase of SN~2005hk to roughly match our B band model light curves in order to better compare the light curve shapes.
Previous work has shown that the too fast decline in the redder bands is, at least to a degree, independent of the overall luminosity (see, e.g., \citealt[Fig.~11]{lach2022b}).
As can be seen in Fig.~\ref{fig:lightcurves}, the CO deflagration model struggles to fully reproduce the width of the observed light curve and declines too fast, particularly in the red bands.
In comparison, the tECSN model has a decline rate much closer to the observed values, particularly for the red bands.
This slower decline is, at least partially, also caused by the lower $M({^{56}}\mathrm{Ni})/M_\mathrm{ej}$ fraction found in the tECSN model compared to the CO deflagration model (see \citealt[Fig.~17]{lach2022b} and Sect.~6.5 therein).
Looking at the colors, one can see that the \modst{} model almost perfectly reproduces the colors of SN~2005hk.
We note that toward later epochs that the approximate non-LTE treatment of \textsc{Artis} becomes less accurate, particularly for NIR wavelengths and a full non-LTE treatment will be necessary (see, e.g., \citealt{collins2025a,holas2025a}).
\citet{callan2023a} showed that non-LTE effects can already be relevant at early phases for deflagration models (also see \citealt{collins2025a} for SN~Ia models in general).
At the same time, multidimensional effects have been shown to play an important role in photospheric-phase RT calculations (\citealt{pollin2024a}), which are currently impractical to perform in non-LTE.
While the model predictions may change with full non-LTE treatment, we estimate the significantly different colors of the model are a result of the explosion model properties themselves, rather than the assumptions of the RT.
The addition of Ti and Cr in the tECSN explosion mechanism seems overall beneficial in producing colors close to those observed in SN~Iax, at least without the inclusion of the bound remnant.\footnote{We only show SN~2005hk here, but the fast decline of CO deflagration compared to SN~Iax observations also holds for other objects and models, as well as the disagreement with observed colors; see \citet[Fig.~11]{lach2022b}.}

Lastly, we briefly mention viewing angle effects.
As mentioned in Sect.~\ref{sec:nucleo}, the ejecta of both models are reasonably well mixed.
The \modlach{} model exhibits a slightly more even mixing, whereas the \modst{} model shows a slightly bipolar structure.
This difference is also reflected in the viewing angle dispersion, illustrated in Fig.~\ref{fig:lightcurves_va}, that is, the tECSN model shows a noticeably larger viewing angle dispersion.
Nonetheless, for any viewing angle, the tECSN model is still characterized by a slower decline in the redder bands.
As our main conclusion appears to be robust against viewing angle variation, we defer  a detailed investigation of viewing angle effects to future work to keep the present discussion focused.

\subsection{Spectra}\label{sec:early_spec}
\begin{figure*}
    \centering
    \includegraphics[]{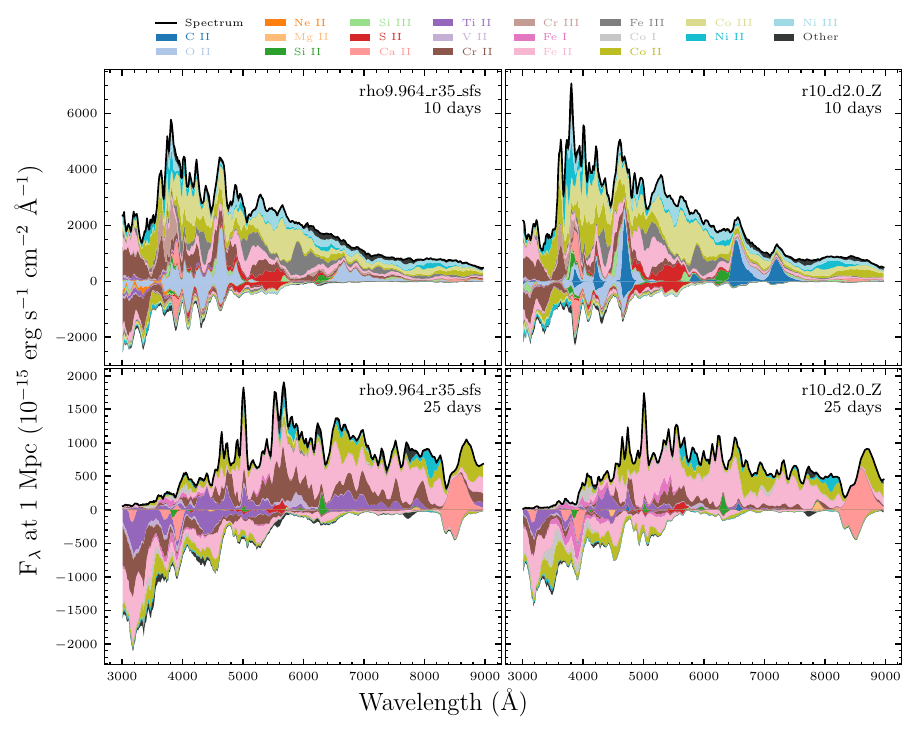}
    \caption{Spectral energy decomposition plots of both explosion models at $10$ and $25\,\mathrm{days}$ after the explosion. Here, we only color coded the most significant contribution to the spectral emission and absorption flux.}
    \label{fig:sdec_early}
\end{figure*}
As already discussed in Sect.~\ref{sec:early_lc}, one of the main differences between the tECSN and CO deflagration explosion models is the redder color of tECSNe as well as a slower decline rate, particularly at longer wavelengths.
Here, we take a closer look at the spectral composition and what elements cause this difference (for a full time series of the early-time spectra, see Fig.~\ref{fig:spec_time_early}).

Looking at the spectra at $10\,\mathrm{days}$ after the explosion in Fig.~\ref{fig:sdec_early}, we find that although both spectra look similar, there are some notable differences.
First, the \modlach{} model shows several clear C lines (in particular $\lambda6580$ and $\lambda7234$) which are completely absent from the \modst{} model.
This is the first key difference in the spectral appearance of these two explosion mechanisms.
As the tECSN scenario can only occur in degenerate ONe cores, they are naturally deficient of C (which also does not get produced during the explosion) and therefore this explosion mechanism will not cause any C features.
However, this only pertains to the explosion mechanism itself, not signatures of the progenitor.
For example, an ONe tECSN progenitor might be surrounded by a reasonably thick shell of residual C left over from the preceding stellar evolution \citep[e.g.,][]{schwab2019b}, which will get mixed into the ejecta, thereby potentially contaminating the spectral appearance with C features.

To estimate the potential impact of residual C, we considered the two \modst{} model variations, the \modst{}\_0.01C and \modst{}\_0.01C\_0.5Cshell model.
We find that when ignoring the individual C features, the resulting RT calculations yield very similar results compared to the original tECSN model.
For example, the light curves of the \modst\_0.01C model are mostly identical, whereas the \modst\_0.01C\_0.5Cshell model exhibits a $0.3\,\mathrm{days}$ earlier B band peak.
Focusing on the difference in the \ion{C}{ii} emission, illustrated in Fig.~\ref{fig:sdec_c_app}, we find that although contributing to the overall spectral appearance the core contamination is not enough to cause clearly visible \ion{C}{ii} lines.
This changes drastically for the C shell, that is the \modst\_0.01C\_0.5Cshell model.
Here, the \ion{C}{ii} lines are comparable to those found in the \modlach{} model and the two spectra become nearly indistinguishable.
Therefore, we conclude that \ion{C}{ii} features are not a reliable signature to distinguish a CO deflagration from a tECSN as they can also be produced by a C shell that can be found in tECSN progenitors as well.
However, our results also show that residual C in the core is unlikely to cause visible C features and a C shell is required to produce these features.

Moving to later epochs, where C features are usually no longer visible, we find that the differences already discussed in Sect.~\ref{sec:early_lc} are also reflected in the spectra.
In the lower half of Fig.~\ref{fig:sdec_early}, where the spectra at around $15\,\mathrm{days}$ after peak are depicted, the impact of Ti and Cr can clearly be seen in the case of the tECSN model.
While the overall features are similar in both spectra --- the C feature is no longer contributing to the spectral appearance --- the \modst{} model shows much stronger absorption at blue wavelengths and enhanced flux in at R band wavelengths, both caused by Ti and Cr.
This causes the difference in color highlighted in Sect.~\ref{sec:early_lc}.
In contrast to the C feature, this difference is much more closely connected to the explosion mechanism.
Whereas the C feature could reasonably be explained by contamination in the progenitor, the enhanced Ti and Cr production appears to be a feature exclusive to tECSNe and cannot easily be explained by the progenitor evolution or solar metallicity.
At this epoch it also becomes apparent that although there are other differences in the elemental composition, these do not significantly impact the overall spectral appearance.
For example, the Mg production in the two explosion channels is quite substantial (the \modlach{} model has a factor $2.8$ times more Mg by mass), yet barely affects the overall spectral appearance; see the Mg emission at around $7900\,$\AA{} in Fig.~\ref{fig:sdec_early}.

\begin{figure*}
    \centering
    \includegraphics[]{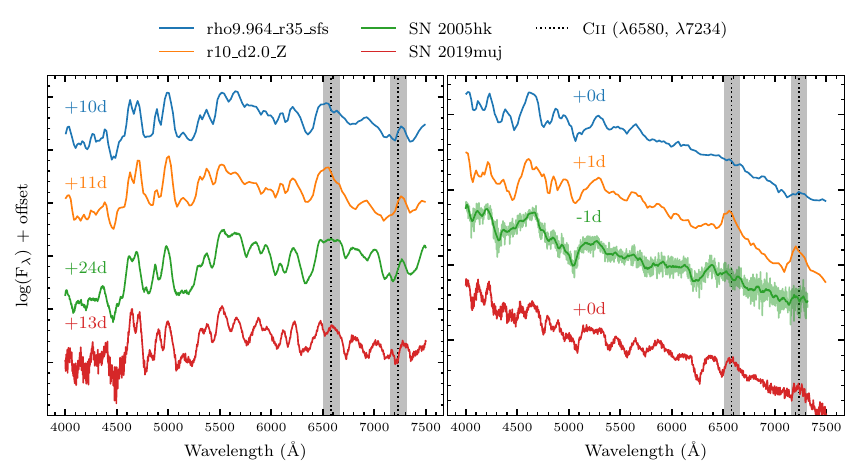}
    \caption{Comparison of our synthetic model spectra to observed spectra of SN~2005hk and SN~2019muj. The data of the observed spectra are from WISeREP \citep{yaron2012a}. Left: Comparison of the spectra after the peak, highlighting similarities and differences in the overall flux. Right: Focus on the \ion{C}{ii} lines around peak brightness.}
    \label{fig:spec_comp}
\end{figure*}
As in the previous section, we also briefly compare our model spectra to observational data, SN~2005hk and SN~2019muj \citep[another well observed SN~Iax;][]{kawabata2021a,barna2021a}, as illustrated in Fig.~\ref{fig:spec_comp}.\footnote{Note that these observed SNe have a luminosity different from our model SNe.}
Here, in the right column of Fig.~\ref{fig:spec_comp}, we briefly discuss the presence of C features in observed spectra; one of the clear differences between our model spectra.
Here, we find that around peak magnitude SN~2005hk does not show a clear \ion{C}{ii} feature\footnote{C features are difficult to detect in observational spectra and can sometimes be hidden by other elements. Therefore the absence of P-Cygni profiles at, for example, \ion{C}{ii} lines is not definitive evidence for the absence of C in the ejecta.} and is more similar to our \modst{} model.
In contrast, SN~2019muj exhibits two P-Cygni profiles at the wavelengths expected for \ion{C}{ii} features, suggesting that C is present in the ejecta of SN~2019muj that the CO deflagration model can easily explain.
We emphasize that C features can also be produced by residual C around the ONe core, as discussed previously.
Consequently, although a clear difference in the explosion mechanisms and a feature that is not consistently observed for all SN~Iax, it is not a feature that can exclusively be explained by either one or the other explosion mechanism. 

In the left column, we compare our spectra at around $10\,\mathrm{days}$ after peak B band brightness, to observations at phases with comparable features.
Both model spectra resemble prototypical SN~Iax in terms of broad spectral appearance and it is likely that they would be placed in this classification (not counting scenarios where the influence of the progenitor significantly impacts the observational signature such as a H envelope).
SN~2005hk is a very close match for both models, although the overall flux profile more closely resembles, as indicated by the light curves, that of \modst{}.
Here, the most striking difference is the plateau around $6600\,$\AA{} that neither of our models exhibit, although in case of the \modst{} model the additional flux from Ti and Cr certainly produces a more similar shape.
SN~2019muj in comparison shows a more distinct set of spectral features, as well as many lines that are not visible in our models.
However, the overall spectral appearance is still similar to that of our models.
Overall, we find that although both models resemble the spectral appearance of observed SN~Iax, the tECSN model can more closely match the overall flux profile due to the additional flux provided by Ti and Cr.

\section{Late-time observables}\label{sec:late}
In this section we discuss the spectral appearance of our models at later phases, around $60$ to $180\,\mathrm{days}$ after the explosion, commonly referred to as the nebular phase.
Here we focus on the NIR and mid-infrared (MIR), particular wavelengths between $6\,\mu\mathrm{m}$ and $20\,\mu\mathrm{m}$, as these wavelengths cover the \ion{Ne}{ii} emission line, which could potentially distinguish CO and ONe deflagrations since Ne is not abundantly produced in CO deflagrations.
It should be mentioned that, as previously discussed, both explosion models leave behind a massive central remnant which we ignore in this work.
The remnant is shown to be important to both the luminosity and the spectra predicted at early-times \citep{callan2024a} and it is likely to also have an impact on the late phase spectra.
In particular, models of pure deflagrations predict that the bound remnant retains substantial amounts of ${^{56}}\mathrm{Ni}$, providing a power source that could potentially drive a steady wind \citep{foley2016a,shen2017a}.
It has been suggested that such a wind may contribute to late-time spectral components and result in low-velocity emission features in observed SNe~Iax \citep{foley2016a,kawabata2018a}.
We leave an investigation of the impact of the bound remnant for future work.

\begin{figure*}
    \centering
    \includegraphics[]{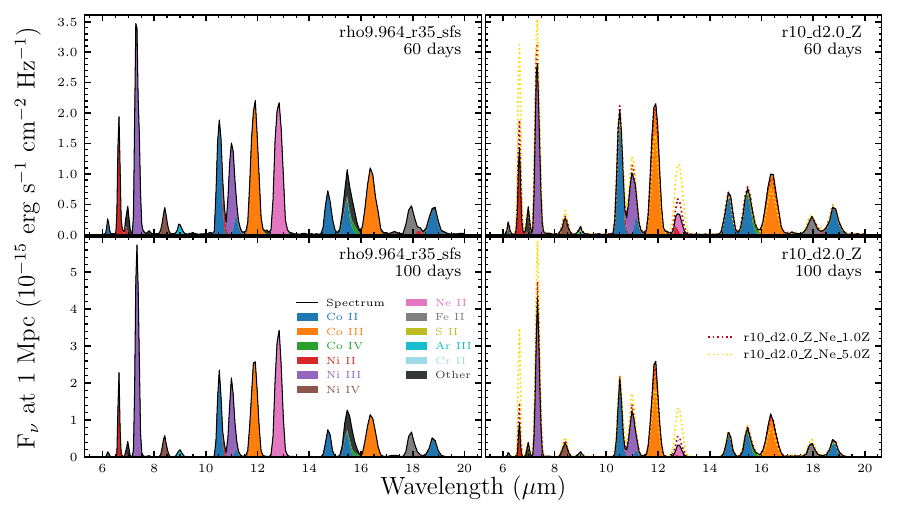}
    \caption{Late-time emission spectra of both explosion models at $60$ and $100\,\mathrm{days}$ after the explosion colored by the last emission type. Here we focus mostly on the \ion{Ne}{ii} feature at $12.8\,\mu\mathrm{m}$.}
    \label{fig:emission}
\end{figure*}
In Fig.~\ref{fig:emission} we illustrate the MIR spectra with the contributions to the emission of different species highlighted (based on the last emission) for both models at two phases (for the full spectral timeseries, see Fig.~\ref{fig:spec_time_late}), at $60$ and $100\,\mathrm{days}$ after the explosion.
Following the pattern found in the previous sections, both explosion models are remarkably similar in their overall appearance, yet some key difference can be spotted, potentially revealing signatures of the explosion mechanism.
First, we focus on the \ion{Ne}{ii} line, as Ne is the dominant element that sets both progenitors apart (in addition to C, see Sect.~\ref{sec:early_spec}).
It becomes immediately evident that this line is significantly stronger in the tECSN case compared to the CO deflagration.
This is predominantly due to the fact that the \modst{} model contains an order of magnitude more Ne left over from the progenitor, whereas any Ne in the \modlach{} model needs to be produced during the explosion --- a relatively small amount in this case.
It can now be argued that the CO WD progenitor could contain some Ne as well as part of its metallicity ($X_\odot({^{22}}\mathrm{Ne)\approx0.02}$, \citealt{asplund2009a}), which we address with the \modlach{}\_Ne\_1.0Z and \modlach{}\_Ne\_5.0Z model variations.
In the $1\times Z_\odot$ case, the spectral appearance barely changes and the strength of the \ion{Ne}{ii} line exhibits only a moderate increase.
Even in the \modlach{}\_Ne\_5.0Z case, the exceptionally strong \ion{Ne}{ii} line observed in the tECSN model cannot be reproduced.
However, we note that in this model variation, the reduction in ${^{56}}\mathrm{Ni}$ production starts to significantly impact the total luminosity and consequently overall ionization state, as can be seen in, for example, \ion{Ni}{ii} and \ion{Ni}{iii} emission lines.
For this reason, the comparison with the \modlach{}\_Ne\_5.0Z model should be taken with some care and future studies investigating this feature should consider models of comparable luminosity.

\begin{figure}
    \centering
    \includegraphics[]{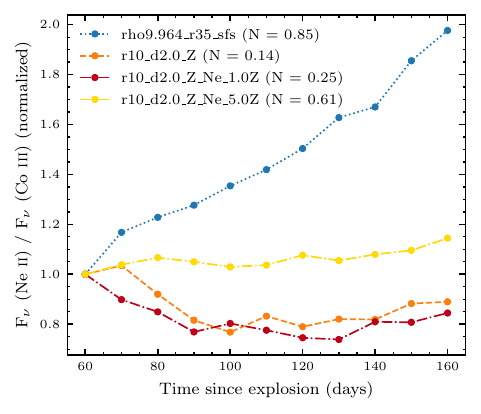}
    \caption{Evolution of the (normalized) flux ration between the peak flux in the \ion{Ne}{ii} ($12.8\,\mu\mathrm{m}$) and \ion{Co}{iii} ($11.89\,\mu\mathrm{m}$) over time.}
    \label{fig:line_ratio}
\end{figure}
In addition to the total line strength, the different explosion channels also exhibit a different time evolution of the line strength.
To illustrate this, in Fig.~\ref{fig:line_ratio} we show the relative line strength (compared to the \ion{Co}{III} line at around $12\,\mu\mathrm{m}$), normalized to the initial ratio.\footnote{This ratio also depends on the evolution of the \ion{Co}{iii} flux, which is not constant in both models. However, this trend can also be reproduced when looking at the total \ion{Ne}{ii} flux.}
Here, it can be clearly seen that while the line ratio in the CO deflagration model remains roughly the same, it doubles in the tECSN case.
Even with $5\times Z_\odot$ worth of ${^{22}}\mathrm{Ne}$, the \modlach{}\_Ne\_5.0Z model cannot reproduce this line ratio evolution.

\begin{figure}
    \centering
    \includegraphics[]{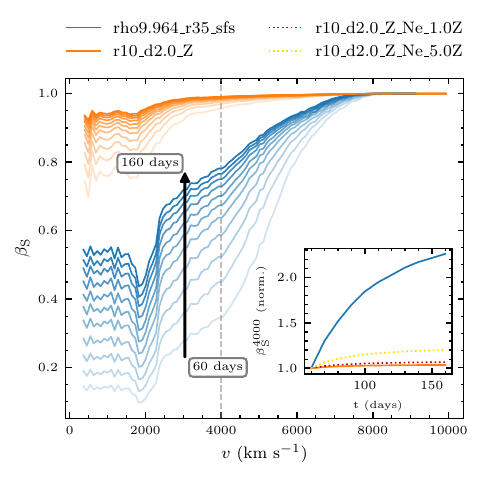}
    \caption{Sobolev escape probability, $\beta_\mathrm{S}$, for the \ion{Ne}{ii} transition for the \modst{} and \modlach{} models. The x-axis shows the initial velocity coordinate of each radial shell of the input model. We show $\beta_\mathrm{S}$ for various epochs between $60$ and $160$ days with a $10$-day spacing. The escape probability increases with time as indicated by the upward arrow. In the inset plot, we show the normalized value of $\beta_\mathrm{S}$ at the $4000\,\mathrm{km}\,\mathrm{s}^{-1}$ shell over time. We also include the \modlach{} model variations.}
    \label{fig:self_abs}
\end{figure}
The reason for this significantly different behavior lies in the different ejecta structure between the two models.
Importantly, the more compact (i.e., higher density) ejecta structure of the \modst{} model, in combination with the overall higher mass fraction of Ne (even compared to the \modlach{}\_Ne\_5.0Z the Ne mass fraction in the \modst{} model is still a factor of three higher) leads to self-absorption of the \ion{Ne}{ii} line.
This self-absorption retains much of the emission originating from the \ion{Ne}{ii} line transition.
As the ejecta expand (both models expand with comparable velocities) and density decreases, an increasing amount of photons from this transition escape, which leads to a strengthening of this line.
In contrast, in the \modlach{} model, the initial ejecta densities and Ne mass fractions are low enough that the fraction of self-absorption is rather small.
To illustrate this, we computed the Sobolev escape probability $\beta_\mathrm{S}$, which is defined as
\begin{gather}
    \beta^{ij}_\mathrm{S} = \frac{1 - \exp\left(-\tau^{ij}_\mathrm{S}\right)}{\tau^{ij}_\mathrm{S}},\\
    \tau^{ij}_\mathrm{S} = \frac{A_{ij}\lambda_{ij}^3}{8\pi}\left(\frac{g_i}{g_j}n_j -n_i\right) t_\mathrm{E},
\end{gather}
following the formalism of, for example, \citet{jeffery1995a,lucy2002a,noebauer2019a,jerkstrand2025a}.
Here, $\tau_\mathrm{S}$ is the Sobolev optical depth, $A$ is the Einstein coefficient, $n$ is the level population, $g$ is the statistical weight, and $t_\mathrm{E}$ is the time since explosion.
The index $ij$ indicates the levels involved in the transition, in the case of the \ion{Ne}{ii} line $i=2\mathrm{s}^2\,2\mathrm{p}^5\,2\mathrm{P}^\mathrm{o}\,(1/2)$ and $j=2\mathrm{s}^2\,2\mathrm{p}^5\,2\mathrm{P}^\mathrm{o}\,(3/2)$.
Fig.~\ref{fig:self_abs} shows $\beta_\mathrm{S}$ for the tECSN and CO deflagration model for each radial shell at various epochs between $60$ and $160$ days.
Here, the substantial increase probability in case of the \modst{} model becomes evident, whereas the \modlach{} model exhibits a rather high escape probability throughout all epochs.
If we follow $\beta_\mathrm{S}$ at around $4000\,\mathrm{km}\,\mathrm{s}^{-1}$,\footnote{The FWHM of the \ion{Ne}{ii} line at $60$ days is roughly $7000\,\mathrm{km}\,\mathrm{s}^{-1}$.} illustrated in the inset plot in Fig.~\ref{fig:self_abs}, we find that the escape probability in the \modst{} model more than doubles, which qualitatively tracks the increase in the \ion{Ne}{ii} line strength increase shown in Fig.~\ref{fig:line_ratio}.
Similarly, the \modlach{} model and its variations show a more or less constant $\beta_\mathrm{S}$, also tracking their respective line strength evolution.

Within the framework investigated here, the strengthening of the \ion{Ne}{ii} line over time appears to indeed be a characteristic feature of the tECSN explosion channel, as both the high density of the tECSN ejecta and the substantial Ne abundance are required.
Simply increasing the Ne content in the CO deflagration model does not seem to lead to the same effect.
Although it remains to be determined how well this holds for other tECSN and CO deflagration models, the feature appears ot be robust for models of comparable luminosity; for example the N1def model of \citet{fink2014a} leads to the same conclusion as the \modlach{} model.
It also remains to be seen whether pure deflagrations in hybrid CONe WDs \citep[e.g.,][]{kromer2015a} or low-density ONe WDs \citep[e.g.,][]{marquardt2015a} could produce similar behavior.

Another interesting difference is the strength of the Ni lines.
In general, the Ni lines (\ion{Ni}{ii-iv} in Fig.~\ref{fig:emission}) are noticeably stronger in the tECSN model.
This appears to be a result of the overall larger stable Ni mass in the ejecta of the \modst{} model, which has around $40\,\%$ more stable Ni than the \modlach{} model by mass (see Sect.~\ref{sec:yields}); another trace of the explosion mechanism.

\begin{figure*}
    \centering
    \includegraphics[]{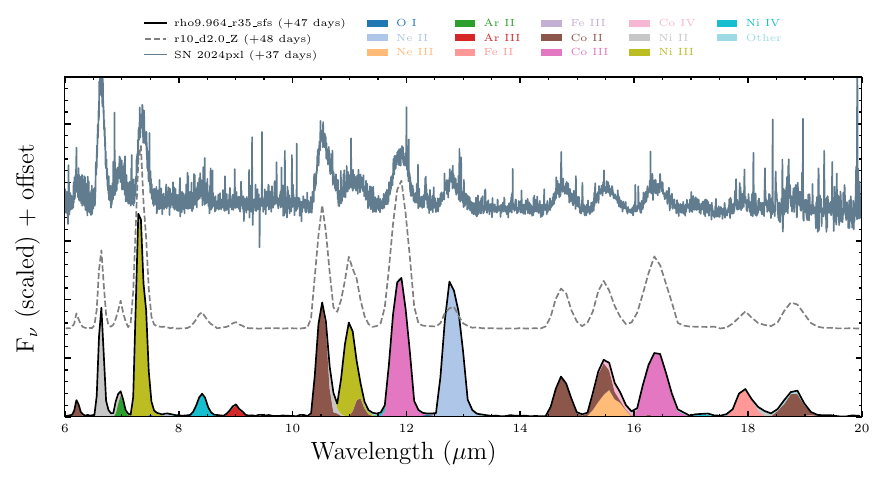}
    \caption{Comparison of our late-time model spectra to a NIR+MIR spectrum of SN~2024pxl (data taken from WISeREP; \citealt{yaron2012a}). We illustrate the spectral energy decomposition only for the \modst{} model.}
    \label{fig:jwst}
\end{figure*}
We also briefly compare our late-time spectra to observations, comparing the previously discussed explosion mechanism traces.
In Fig.~\ref{fig:jwst}, we compare a MIR spectrum of SN~2024pxl \citep{kwok2025a} $37\,\mathrm{days}$ after peak brightness to our model spectra $57\,\mathrm{days}$ after the explosion.
Here, in terms of overall features, we find that our models are again rather similar to those of SN~2024pxl, with two notable exceptions.
As discussed previously, the \modlach{} model exhibits only a weak \ion{Ne}{ii} line, whereas this feature is significantly more pronounced in the \modst{} case.
Comparing this to the observed spectrum, one can see that here the \ion{Ne}{ii} strength is comparable to that of the \ion{Co}{iii} line, which is similar to our tECSN model (in particular if one extrapolates the line ratio evolution; see Fig.~\ref{fig:line_ratio}).
It should be noted that even with a solar amount of Ne in the CO deflagration progenitor, this result does not change substantially, see Figs.~\ref{fig:emission} and \ref{fig:line_ratio}.
Furthermore, we find that both our models exhibit an inverse \ion{Ni}{ii} ($6.6\,\mu\mathrm{m}$) to \ion{Ni}{iii} ($7.4\,\mu\mathrm{m}$) line strength compared to the one found in the observed spectrum.
As discussed before, although indicative of a high Ne content, the absolute line strength should be evaluated in combination with the line strength evolution, revealing details of the internal ejecta structure and potentially the explosion mechanism.
Unfortunately, at the time of writing, there exist no observed timeseries of the \ion{Ne}{ii} line of a SN~Iax-like transient with sufficient epoch coverage.

As noted in the comparison above, our models appear to be overall more highly ionized; an effect that could at least partially be attributed to the difference in luminosity.
SN~2024pxl is around $0.7\,\mathrm{mag}$ fainter \citep{singh2025a} than our model spectra, suggesting an overall lower ${^{56}}$Ni mass and consequently different luminosity.
We emphasize that detailed comparisons with observations are not the focus here and an in-depth investigation of the ionization states relative to SN~2024pxl are outside the scope of this work.
Our comparisons, however, demonstrate the power of combining MIR observations and simulations and highlight the need for MIR observations.

In summary we find for our two models, the \ion{Ne}{ii} emission line, particular its strengthening over time, is a strong indicator for the tECSN explosion mechanism. 
Both the high Ne abundances and compact ejecta found in tECSNe are needed to produce the characteristic line strength evolution found here.
In the future, pure deflagrations in low-density ONe WDs and hybrid CONe WDs should also be considered in future studies to evaluate to what extent the high-density environment of tECSNe can be connected to this mechanism.

\section{Impact of a hydrogen envelope}\label{sec:progenitor}

\begin{figure}
    \centering
    \includegraphics[]{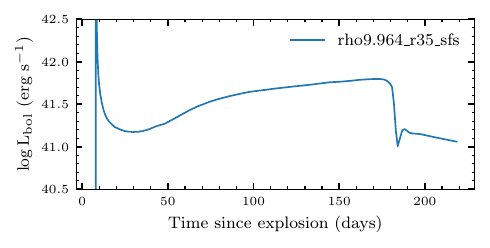}
    \caption{Synthetic light curve of the \modst{} model generated with \textsc{Stella}, assuming that the explosion takes place within an sAGB star H envelope.}
    \label{fig:stella}
\end{figure}
Finally, we also briefly consider the impact of a possible H envelope on the resulting synthetic light curve.
Fig.~\ref{fig:stella} illustrates the bolometric light curve obtained with \textsc{Stella}.
Broadly speaking, the inclusion of an H envelope results in a SN~IIP-like appearance, which is the expected outcome \citep{kozyreva2024a}.
The light curve plateau is driven by H recombination, the same as in a regular, core-collapse SN~IIP.
The tail after the plateau is then powered by radioactive decay of ${^{56}}$Ni and ${^{56}}$Co.
We note that the dip at the beginning of the radioactive tail is caused by a numerical viscosity.
The light curve of our \modst{} model is rather faint and exhibits a particularly long plateau duration, around $170$\,days, caused by the rather low kinetic energy of the ejecta.
As far as we can tell, there exist no observed SNe~IIP that exhibits such a long plateau duration.
For comparison, the e88 cECSN model discussed by \citet{kozyreva2021a} contains over one order of magnitude less ${^{56}}$Ni than our model.
Its plateau duration is only around $120\,\mathrm{days}$, well within the range of observed SN~IIP plateau durations.
Given that the single star progenitor is the least likely tECSN formation channel by a large margin and the overall low expected occurrence rate of tECSN \citep{jones2019a}, the lack of an observed counterpart is not entirely unexpected.
However, it is not excluded that other tECSN models, particularly those with higher kinetic ejecta energies (see \citealt{holas2026a}) could produce shorter plateaus.
As discussed by \citet{kozyreva2024a}, spectra after the plateau phase will most likely enable the differentiation between thermonuclear and core-collapse SNe~IIP.
We leave an investigation of the spectral appearance of single star tECSNe for future work.

\section{Conclusions}\label{sec:conclusion}

In this work, we have investigated the observational appearance of a tECSN explosion, focusing on the binary star and AIC progenitor channel, that is, without the presence of an H envelope.
For comparison, we also computed synthetic observables of a CO deflagration model, the \modlach{} model of \citet{lach2022b}, as this model is characterized by a similar nuclear energy release and overall ejecta composition, but it describes a more conventional thermonuclear SN progenitor channel.
Throughout this work, we also compare our synthetic observables with observations, highlighting the differences between our models in the context of astronomical transients.

In Sect.~\ref{sec:nucleo} we highlighted the differences in ejecta structure and composition between our tECSN model and the CO deflagration model of \citet{lach2022b}.
Importantly, our \modst{} model produces significantly more elements in the range of Sc to Mn and is characterized by a slightly more compact ejecta structure.
Both models exhibit well-mixed ejecta overall, typical for pure deflagrations, and therefore viewing angle effects play a subdominant role.
In the case of the tECSN models, we find that the ejecta are composed of a large variety of radioactive isotopes by a substantial fraction.
While the \modlach{} model (and other thermonuclear SNe models such as double-detonations; e.g., \citealt{shen2018a,pakmor2022a}, for that matter) is dominated by ${^{56}}$Ni and ${^{57}}$Ni
The tECSN model also shows noticeable contributions of, for example, ${^{55}}$Co and ${^{66}}$Ni.
Consequently, RT calculations of tECSNe models need to consider these isotopes as well in order to accurately estimate the energy deposition rate.

In Sect.~\ref{sec:early} we focused on synthetic early-time observations of both models obtained from 3D RT calculations.
We found that both models closely resemble the observational properties of SNe~Iax-like events, in terms of both light curves and spectra.
However, our tECSN model shows an increased flux in the wavelength range of $5500-8000\,$\AA{} compared to the CO deflagration model. 
In particular this is due to the comparably higher Ti and Cr abundance.
Consequentially, the \modst{} model shows a significantly redder V-R color evolution that is closer to what is seen in observed SNe~Iax.
This difference in color evolution appears to be a strong indicator of the tECSN explosion mechanism, as the required Ti and Cr abundances can be traced back to the high-density regime of tECSNe, which cannot easily be created in CO or low-density ONe deflagrations.
A prominent feature in the early spectra of the \modlach{} model are its \ion{C}{ii} P-Cygni features, which can also be seen in some observed SNe~Iax.
We find that this feature can also be reproduced by a C shell around our tECSN ONe WD progenitor as a byproduct of the preceding stellar evolution.
Consequently, the C features cannot be used as a discerning feature between the CO deflagration and tECSN explosion mechanism.

In Sect.~\ref{sec:late} we discussed synthetic late-time observables obtained from 1D spherically averaged version of our models, focusing on NIR and MIR wavelengths.
At these later epochs, our models are again rather similar in their overall appearance.
The most notable difference is the \ion{Ne}{ii} emission line at $12.8\,\mu\mathrm{m}$, which is a dominant emission line in the \modst{} model but only barely visible in the \modlach{} model.
In particular, this line undergoes significant strengthening in the tECSN case, whereas its strength remains more or less constant in the CO deflagration model.
This difference is caused by the relatively high Ne abundance and compact ejecta structure of the tECSN model, leading to an initially high self-absorption rate for photons emitted from the \ion{Ne}{ii} line transition.
Only as the density drops can photons from this transition efficiently escape, leading to a strengthening of the \ion{Ne}{ii} emission line over time.
It is not yet clear if this mechanism is uniquely connected to tECSNe or if other Ne-rich explosions (such as pure deflagrations in low-density ONe WDs or hybrid CONe WDs) could also produce such a feature.
Our experiments indicate that a solar amount of initial Ne in the CO deflagration model does not significantly impact the appearance of the \ion{Ne}{ii} line.
Within the framework of this study, this strengthening over time appears to be a strong indicator of the tECSN explosion channel.

Furthermore, we compared our models to MIR observations of SN~2024pxl, a recently observed SN~Iax \citep{singh2025a,kwok2025a}.
Overall, we find that our models resemble the observations rather well, although some differences remain.
Notably, SN~2024pxl exhibits a rather strong \ion{Ne}{ii} emission line, indicating a high Ne content in the ejecta.
However, due to the lack of MIR observations at later epochs, we were unable to investigate if the strengthening of the \ion{Ne}{ii} line can also be found in observed SNe~Iax-like objects.
This comparison nonetheless highlights the need for future MIR observations of SNe~Iax covering multiple phases, as these nebular spectra promise to reveal deeper insights into the progenitor and explosion mechanism of SNe~Iax.

Lastly, in Sect.~\ref{sec:progenitor}, we also presented the impact of a potential H envelope on the resulting tECSN light curve.
As expected from previous research \citep{kozyreva2024a}, the inclusion of such an envelope creates an SN~IIP-like light curve.
However, due to the comparably high ${^{56}}$Ni content, the plateau duration is extremely long with around $170\,\mathrm{days}$ --- a duration that does not match any observed SN.
This is in line with the expectation that the single star progenitor channel is the least likely tECSN formation channel.

This study has presented first-of-their-kind synthetic observables for a tECSN. 
The main motivation of this work was to determine whether tECSNe could produce observables consistent with those of already observed transients, which would indicate that this explosion channel might actually occur in nature.
Our results show that the predicted observables are very similar to several observed events, and we did not find features that are in tension with existing transient observations.
This provides support for the possibility that tECSNe occur in nature and might have already been observed.
In particular, if such an event has already been observed, it was most likely classified as a SN~Iax-like event (assuming the absence of a H envelope).
Differentiating these events from other progenitor channels, for example a CO deflagration, appears to be somewhat nontrivial, since there are few direct signatures.
Both the V-R color and the \ion{Ne}{ii} line strength evolution seem to be promising signatures of a tECSN explosion, yet it is not clear if these features can be used as a unique identifier.
Other, less direct measurements can potentially also be considered, such as the extremely high $X({^{60}}\mathrm{Fe})/X({^{26}}\mathrm{Fe)}$ ratio found in tECSNe compared to CO deflagrations (see, e.g., \citealt{jones2019a}).
Our work only considers two particular realizations of a tECSN and CO deflagration.
However, as shown by \citet{lach2022b} and \citet{holas2026a}, both explosion channels can yield a large variety of ejecta properties (primarily the total ejecta mass) depending on the initial conditions such as the ignition.
Therefore, our investigations are somewhat limited in their predictive power in terms of explaining actual observations.
Nonetheless, the results of this study are indicative of the overall similar appearance of tECSNe and CO deflagrations, and they indicate that late-time, nebular observations are likely needed to differentiate these explosion mechanisms.

Finally, it should be noted that the particular model realizations here likely represent the bright end of deflagrations in ONe WDs and the faint end of deflagrations in CO WDs.
Therefore, it is conceivable that both explosion channels simultaneously contribute to the observed class of SN~Iax events.
In particular considering that fainter SNe~Iax require lower $M({^{56}}\mathrm{Ni})/M_\mathrm{ej}$ ratios that can be achieved with current CO deflagration models \citep{lach2022b}, the tECSN explosion channel, with its lower radioactive material fraction might offer a convenient alternative for explaining the fainter end of SNe~Iax.
At the same time, neither of the two models discussed here provide a good explanation for LP-40 stars, as their bound remnants are significantly more massive than the low-mass remnants inferred for this class of hyper-velocity star that has been linked to SNe~Iax \citep{bhat2026a}, irrespective of the insufficient kick velocity.
Overall, this underlines that various aspects of both explosion channels remain unexplored and further work is required to holistically constrain their place among observed thermonuclear transients.

\section*{Data availability}
Synthetic light curves and spectra of all models discussed in this work are available on Zenodo.\footnote{\url{https://doi.org/10.5281/zenodo.20445112}}

\begin{acknowledgements}
The authors thank Stefan Taubenberger for his help in classifying the observational characteristics of the synthetic spectra.

A.H. is a fellow of the International Max Planck Research School for Astronomy and Cosmic Physics at the University of Heidelberg (IMPRS-HD) and acknowledges financial support from IMPRS-HD.
CEC is funded by the European Union’s Horizon Europe
research and innovation programme under the Marie Skłodowska-Curie grant
agreement No.~101152610.

This work received support from the European Research Council (ERC) under the European Union’s Horizon 2020 research and innovation programme under grant agreement No.\ 759253 and 945806, the Klaus Tschira Foundation, and the High Performance and Cloud Computing Group at the Zentrum f{\"u}r Datenverarbeitung of the University of T{\"u}bingen, the state of Baden-W{\"u}rttemberg through bwHPC and the German Research Foundation (DFG) through grant no INST 37/935-1 FUGG.
The authors gratefully acknowledge the Gauss Centre for Supercomputing e.V. (www.gauss-centre.eu) for funding this project by providing computing time through the John von Neumann Institute for Computing (NIC) on the GCS Supercomputer JUWELS at Jülich Supercomputing Centre (JSC).
This work used the DiRAC Memory Intensive service (Cosma8) at Durham University, managed by the Institute for Computational Cosmology on behalf of the STFC DiRAC HPC Facility (www.dirac.ac.uk). The DiRAC service at Durham was funded by BEIS, UKRI and STFC capital funding, Durham University and STFC operations grants.
The authors acknowledge support by the state of Baden-Württemberg through bwHPC
and the German Research Foundation (DFG) through grant INST 35/1597-1 FUGG.
This work was supported by the Deutsche Forschungsgemeinschaft (DFG, German Research Foundation) -- RO 3676/7-1, project number 537700965,
and by the European Union (ERC, ExCEED, project number 101096243). Views and opinions expressed are, however, those of the authors only and do not necessarily reflect those of the European Union or the European Research Council Executive Agency. Neither the European Union nor the granting authority can be held responsible for them.
This work was performed on the HoreKa supercomputer funded by the Ministry of Science, Research and the Arts Baden-Württemberg and by the Federal Ministry of Education and Research.
F.P.C. and S.A.S. acknowledge funding from UKRI STFC grant ST/X00094X/1. 
This work was supported by the U.S. Department of Energy through the Los Alamos National Laboratory. Los Alamos National Laboratory is operated by Triad National Security, LLC, for the National Nuclear Security Administration of U.S. Department of Energy (Contract No. 89233218CNA000001).
\end{acknowledgements}

\bibliographystyle{aa}
\bibliography{references}

\onecolumn
\begin{appendix}

\section{Additional figures}
\begin{figure*}[!h]
    \centering
    \includegraphics[width=0.9\textwidth]{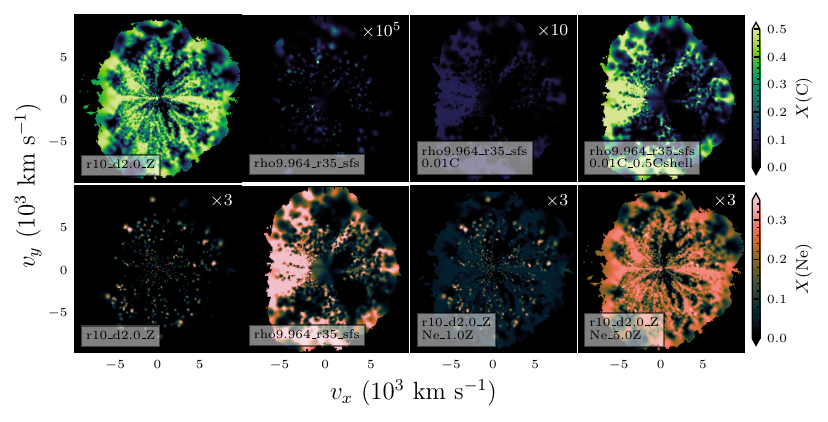}
    \caption{Slices of the mapped C and Ne abundances of both the original explosion models, as well as the C and Ne model variations at $100\,\mathrm{s}$ after the explosion. Note that for several models we have enhanced the respective mass fraction by the factor indicated in the top right corner, for better visibility. The depicted slices are at a resolution of $200^3$, the actual input for model for \textsc{Artis} has a resolution of $50^3$ in 3D and $100$ radially averaged shells in 1D.}
    \label{fig:artis_c_ne_app}
\end{figure*}

In this appendix we include several figures supplementing the main text.
Here, we how the C and Ne composition of the ejecta of both the investigated models and their variations in Fig.~\ref{fig:artis_c_ne_app}.
In Fig.~\ref{fig:sdec_c_app} we illustrate the impact of C left over by the progenitor on the spectral appearance.
Lastly, full spectral time series of both early- and late-time spectra are shown in Fig.~\ref{fig:spec_time_early} and Fig.~\ref{fig:spec_time_late}, respectively.

\begin{figure*}[!h]
    \centering
    \includegraphics[width=0.9\textwidth]{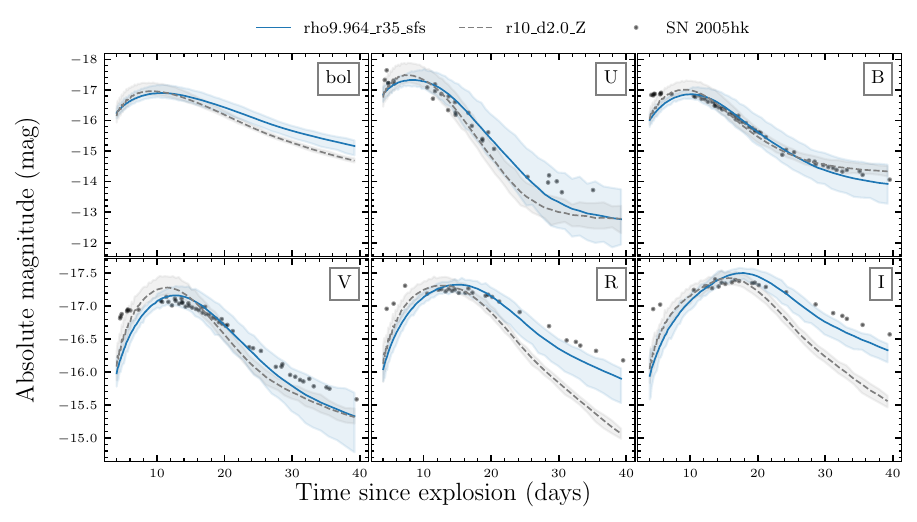}
    \caption{Synthetic early-time light curves of both explosion models. The shaded regions indicate the viewing angle dispersion, and the solid line describes the spherically averaged quantity that we discuss in Sect.~\ref{sec:early_lc}. We also show SN~2005hk with adjusted magnitude and phase, same as in Fig.~\ref{fig:lightcurves}.}
    \label{fig:lightcurves_va}
\end{figure*}

\begin{figure*}[!h]
    \centering
    \includegraphics[]{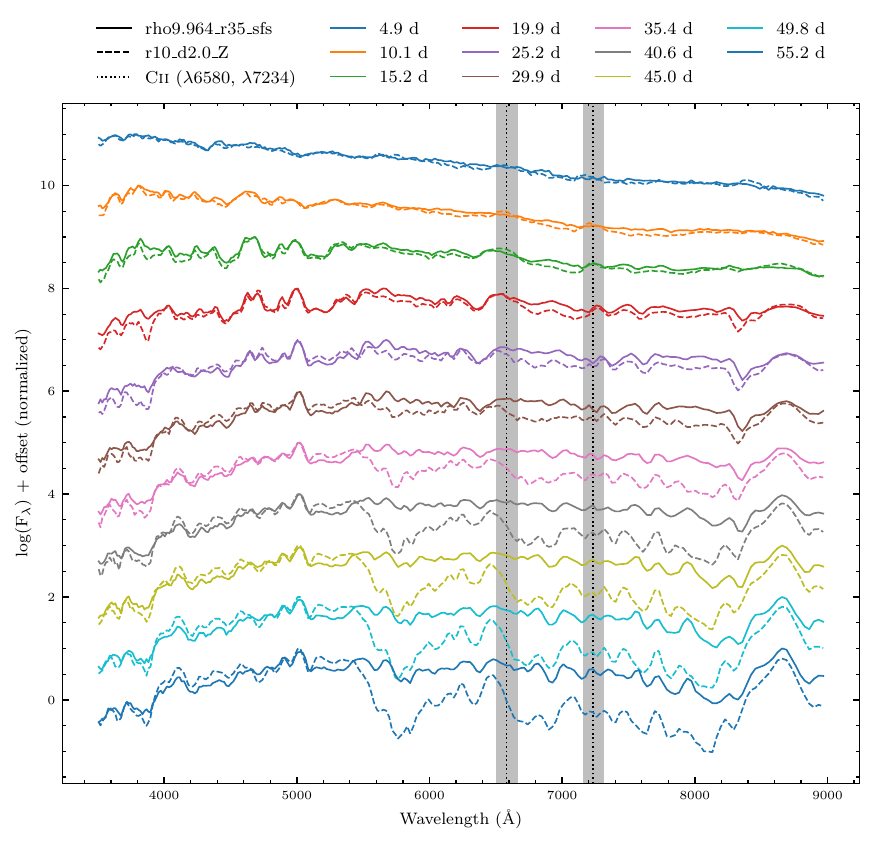}
    \caption{Spectral time series of both explosion models at early-times. The times in the legend are with respect to the explosion. Here, we only show the angle-averaged spectra of the 3D RT simulation.}
    \label{fig:spec_time_early}
\end{figure*}

\begin{figure*}
    \centering
    \includegraphics[]{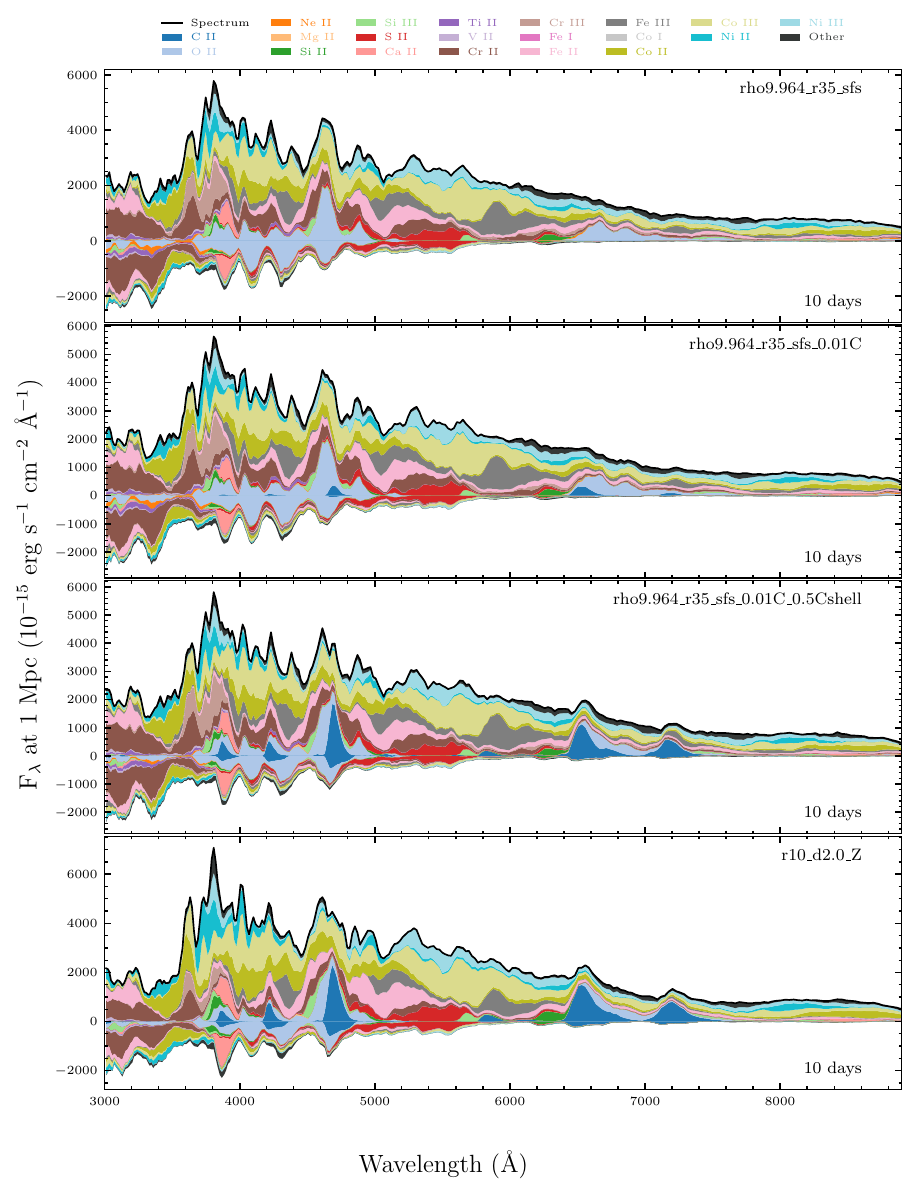}
    \caption{Spectral energy decomposition plots of both explosion models at 10 days after the explosion, as well as the C variations of the \modst{} model. Here, we only color code the most significant contribution to the spectral emission and absorption flux.}\label{fig:sdec_c_app}
\end{figure*}

\begin{figure*}[!h]
    \centering
    \includegraphics[]{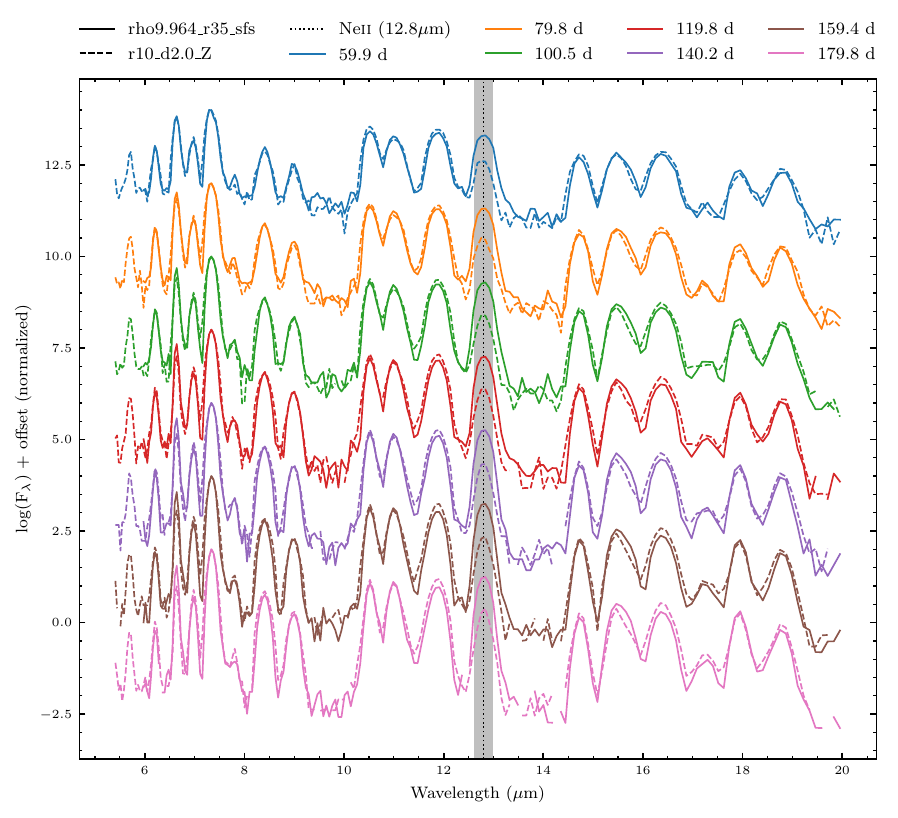}
    \caption{Spectral time series of both explosion models at late times. The times in the legend are with respect to the explosion. Here, we focus on the MIR range, highlighting the different evolution of the \ion{Ne}{ii} line strength.}
    \label{fig:spec_time_late}
\end{figure*}
\end{appendix}

\end{document}